\documentstyle[12pt,epsf]{article}
\newlength{\pubnumber} 

\catcode`\@=11
\@addtoreset{equation}{section}

\def\section{\@startsection{section}{1}{\z@}{3.5ex plus 1ex minus .2ex}
 {2.3ex plus .2ex}{\large\bf}}
\def\subsection{\@startsection{subsection}{2}{\z@}{2.3ex plus .2ex}
 {2.3ex plus .2ex}{\bf}}

\newfont{\mbm}{msbm10 scaled\magstep1}
\def\bb#1{\hbox{\mbm #1}}

\DeclareMathAlphabet\Scr{U}{rsf}{m}{n}

\newcommand{\cc}[2]{c{#1\atopwithdelims[]#2}}

\newcommand{\nn}{\nonumber}

\input epsf

\begin{document}

\begin{titlepage}
\samepage{
\setcounter{page}{1}
\rightline{OUTP--04--21P}
\rightline{\tt hep-th/0411118}
\rightline{October 2004}
\vfill
\begin{center}
 {\Large \bf Self--Duality and Vacuum Selection\\}
\vfill
\vfill {\large
	 Alon E. Faraggi\footnote{faraggi@thphys.ox.ac.uk}}\\
\vspace{.12in}
{\it  The Rudolf Peierls Centre for Theoretical Physics,\\
University of Oxford,Oxford OX1 3NP, United Kingdom\\}
\vspace{.025in}
\end{center}
\vfill
\begin{abstract}

I propose that self--duality in quantum phase--space provides
the criteria for the selection of the quantum gravity vacuum.
The evidence for this assertion arises from two independent
considerations. 
The first is the phenomenological success of the free fermionic
heterotic--string models, which are constructed in the vicinity of the 
self--dual point under T--duality. The relation between the
free fermionic models and the underlying $Z_2\times Z_2$
toroidal orbifolds is discussed. Recent analysis revealed
that the $Z_2\times Z_2$ free fermionic orbifolds utilize
an asymmetric shift in the reduction to three generations,
which indicates that the untwisted geometrical moduli
are fixed near the self--dual point. The second consideration
arises from the recent formulation of quantum mechanics from an
equivalence postulate and its relation to phase--space duality.
In this context it is demonstrated that the trivial state,
with $V(q)=E=0$, is identified with the self--dual state
under phase--space duality. These observations suggest
a more general mathematical principle in operation. In physical
systems that exhibit a duality structure, the self--dual
states under the given duality transformations correspond to
critical points.
\end{abstract}
\smallskip}
\end{titlepage}

\setcounter{footnote}{0}

\def\beq{\begin{equation}}
\def\eeq{\end{equation}}
\def\beqn{\begin{eqnarray}}
\def\eeqn{\end{eqnarray}}

\def\no{\noindent }
\def\nolabel{\nonumber }
\def\ie{{\it i.e.}}
\def\eg{{\it e.g.}}
\def\half{{\textstyle{1\over 2}}}
\def\third{{\textstyle {1\over3}}}
\def\quarter{{\textstyle {1\over4}}}
\def\sixth{{\textstyle {1\over6}}}
\def\m{{\tt -}}
\def\p{{\tt +}}

\def\Tr{{\rm Tr}\, }
\def\tr{{\rm tr}\, }

\def\slash#1{#1\hskip-6pt/\hskip6pt}
\def\slk{\slash{k}}
\def\GeV{\,{\rm GeV}}
\def\TeV{\,{\rm TeV}}
\def\y{\,{\rm y}}
\def\SM{Standard--Model }
\def\SUSY{supersymmetry }
\def\SSSM{supersymmetric standard model}
\def\vev#1{\left\langle #1\right\rangle}
\def\l{\langle}
\def\r{\rangle}
\def\o#1{\frac{1}{#1}}

\def\Htw{{\tilde H}}
\def\chibar{{\overline{\chi}}}
\def\qbar{{\overline{q}}}
\def\ibar{{\overline{\imath}}}
\def\jbar{{\overline{\jmath}}}
\def\Hbar{{\overline{H}}}
\def\Qbar{{\overline{Q}}}
\def\abar{{\overline{a}}}
\def\alphabar{{\overline{\alpha}}}
\def\betabar{{\overline{\beta}}}
\def\tautwo{{ \tau_2 }}
\def\thetatwo{{ \vartheta_2 }}
\def\thetathree{{ \vartheta_3 }}
\def\thetafour{{ \vartheta_4 }}
\def\ttwo{{\vartheta_2}}
\def\tthree{{\vartheta_3}}
\def\tfour{{\vartheta_4}}
\def\ti{{\vartheta_i}}
\def\tj{{\vartheta_j}}
\def\tk{{\vartheta_k}}
\def\calF{{\cal F}}
\def\smallmatrix#1#2#3#4{{ {{#1}~{#2}\choose{#3}~{#4}} }}
\def\ab{{\alpha\beta}}
\def\Minv{{ (M^{-1}_\ab)_{ij} }}
\def\bone{{\bf 1}}
\def\ii{{(i)}}
\def\V{{\bf V}}
\def\N{{\bf N}}

\def\b{{\bf b}}
\def\S{{\bf S}}
\def\X{{\bf X}}
\def\I{{\bf I}}
\def\mb{{\mathbf b}}
\def\mS{{\mathbf S}}
\def\mX{{\mathbf X}}
\def\mI{{\mathbf I}}
\def\balpha{{\mathbf \alpha}}
\def\bbeta{{\mathbf \beta}}
\def\bgamma{{\mathbf \gamma}}
\def\bxi{{\mathbf \xi}}

\def\t#1#2{{ \Theta\left\lbrack \matrix{ {#1}\cr {#2}\cr }\right\rbrack }}
\def\C#1#2{{ C\left\lbrack \matrix{ {#1}\cr {#2}\cr }\right\rbrack }}
\def\tp#1#2{{ \Theta'\left\lbrack \matrix{ {#1}\cr {#2}\cr }\right\rbrack }}
\def\tpp#1#2{{ \Theta''\left\lbrack \matrix{ {#1}\cr {#2}\cr }\right\rbrack }}
\def\l{\langle}
\def\r{\rangle}


\def\inbar{\,\vrule height1.5ex width.4pt depth0pt}

\def\IC{\relax\hbox{$\inbar\kern-.3em{\rm C}$}}
\def\IQ{\relax\hbox{$\inbar\kern-.3em{\rm Q}$}}
\def\IR{\relax{\rm I\kern-.18em R}}
 \font\cmss=cmss10 \font\cmsss=cmss10 at 7pt
\def\IZ{\relax\ifmmode\mathchoice
 {\hbox{\cmss Z\kern-.4em Z}}{\hbox{\cmss Z\kern-.4em Z}}
 {\lower.9pt\hbox{\cmsss Z\kern-.4em Z}}
 {\lower1.2pt\hbox{\cmsss Z\kern-.4em Z}}\else{\cmss Z\kern-.4em Z}\fi}

\def\AEF{A.E. Faraggi}
\def\NPB#1#2#3{{\it Nucl.\ Phys.}\/ {\bf B#1} (#2) #3}
\def\PLB#1#2#3{{\it Phys.\ Lett.}\/ {\bf B#1} (#2) #3}
\def\PLA#1#2#3{{\it Phys.\ Lett.}\/ {\bf A#1} (#2) #3}
\def\PRD#1#2#3{{\it Phys.\ Rev.}\/ {\bf D#1} (#2) #3}
\def\PRL#1#2#3{{\it Phys.\ Rev.\ Lett.}\/ {\bf #1} (#2) #3}
\def\PRT#1#2#3{{\it Phys.\ Rep.}\/ {\bf#1} (#2) #3}
\def\MODA#1#2#3{{\it Mod.\ Phys.\ Lett.}\/ {\bf A#1} (#2) #3}
\def\IJMP#1#2#3{{\it Int.\ J.\ Mod.\ Phys.}\/ {\bf A#1} (#2) #3}
\def\nuvc#1#2#3{{\it Nuovo Cimento}\/ {\bf #1A} (#2) #3}
\def\RPP#1#2#3{{\it Rept.\ Prog.\ Phys.}\/ {\bf #1} (#2) #3}
\def\APJ#1#2#3{Astrophys. J. {\textbf #1} (#2) #3 }
\def\JHEP#1#2#3{JHEP {\textbf #1} (#2) #3 } 
\def\etal{{\it et al\/}}

\hyphenation{su-per-sym-met-ric non-su-per-sym-met-ric}
\hyphenation{space-time-super-sym-met-ric}
\hyphenation{mod-u-lar mod-u-lar--in-var-i-ant}


\setcounter{footnote}{0}
\section{Introduction}

The study of string theory continues to inspire much interest in theoretical
physics. This arises from the potential of the theory
to probe consistently what may be the necessary ingredients of the
underlying quantum gravity theory, as well as to produce the 
structures that are observed in contemporary high energy experiments.
The study of the theory and its physical implications may be pursued
in two approaches. The first asserts that we must first understand
the mathematical formulation of the theory, and experimental predictions
should only be extracted subsequently. The second aims to develop
a phenomenological approach to string theory. In this context
the first step is to identify string vacua that exhibit viable
phenomenological properties. These vacua should then be utilized
to study the fundamental principles that underly the theory. 
In this article the second approach is discussed.

It is essential to examine the phenomenological approach in view of
the important progress that was made in the fundamental understanding
of string theory over the past decade \cite{reviewMtheory}.
The basic picture which emerged is that all string theories
in ten dimensions, as well as eleven dimensional supergravity are
effective limits of a more fundamental theory,
traditionally dubbed M--theory. 
The question still remains, however, how to utilize these developments
towards phenomenological studies of string theory. From another view,
the new picture of M--theory offers a novel perspective on phenomenological 
studies of string theory, and it is therefore important to
consider what this perspective is. 

The M--theory development indicates that the 
different string theories in ten dimensions as well as the eleven
dimensional supergravity are effective perturbative limits of a more
fundamental theory. The basic formalism of the more fundamental theory is a 
complete mystery at the present time. In this respect the seminal
observation of anomaly cancellation in string theories \cite{greenschwarz}
indicates the consistency of the perturbative
expansion, as well as the inclusion of all the physical degrees of freedom.
On the other hand,
it also indicates that none of the effective string limits can 
fully characterize the true vacuum, which should have some nonperturbative
realization. 

The constructive approach to string phenomenology is to accept 
the limitation of the effective limits. The perturbative string theories
can only probe some of the properties of the true vacuum
but cannot fully characterize it. In this respect it may well be
that some characteristics of the true vacuum can be seen 
in one limit, whereas other characteristics are more readily
gleaned in another limit. For example, one of the appealing properties
of the Standard Model matter spectrum is its embedding in $SO(10)$
spinorial representations. This property of the Standard Model
spectrum can only be realized in the heterotic limit of the 
underlying M--theory. The reason being that only the heterotic
limit gives rise to spinorial representations in the perturbative
spectrum. On the other hand, we know that the perturbative heterotic
string is an expansion in vanishing string coupling and the 
dilaton exhibits a run away behavior in this limit. Furthermore, 
with the advent of the string duality picture, it is reasonable to assume
that in order to stabilize the dilaton at a finite value 
we have to move away from the perturbative heterotic expansion.
This notion is supported by the fact that in the eleven dimensional
limit of M--theory the dilaton becomes the moduli of the 
eleventh dimensions, and hence its stabilization at a finite value
implies the existence of an additional degree of freedom that in this
limit is realized as an extra dimension.

The new picture of M--theory therefore suggests the following
strategy toward its phenomenological studies. The true vacuum
should possess some nonperturbative realization, and the perturbative
string limits can only probe some of its properties. 
Thus, different limits may reveal different properties of the
true vacuum. The development of this methodology over the past
few years has been pursued \cite{oxsp,fknr}
by applying it to the class of $Z_2\times Z_2$ orbifolds
of six dimensional compactified tori. 

{}From the phenomenological point of view, the Standard Model data
strongly hints at the realization of grand unification structures
in nature. This is most appealing in the context of $SO(10)$
Grand Unified Theories,
in which each of the Standard Model generations, including the
right--handed neutrinos, is embedded in a 16 spinorial representation.
If we regard the quantum number of the Standard Model matter states
as experimental observables, as they were in the process of the experimental
discovery of the Standard Model, then the embedding in $SO(10)$ reduces
the number of parameters from 54 to 3. Here 54 counts the number of
multiplets times the number of group factors times the number of families,
whereas 3 counts the number of $SO(10)$ spinorial
representations needed for the embedding. 
Proton longevity implies that the $SO(10)$ unification
can only be realized at a scale which is not much below the string scale.
This picture of high unification scale is also supported by the 
logarithmic running exhibited by the Standard Model parameters.
The logarithmic running is verified in contemporary experiments
in the accessible energy range. The high scale unification paradigm
is also compatible with the experimental data in the gauge and 
heavy generation matter sectors, whereas preservation of the
logarithmic running in the scalar sector requires the introduction
of supersymmetry. 

It should be noted that the $SO(10)$ unification symmetry should be
broken to the Standard Model directly at the string level \cite{nilles1}.
The reason being that this case offers an appealing solution to the
GUT doublet--triplet splitting problem. In this solution the color
triplets are projected out from the effective low energy spectrum,
whereas the doublets remain light. 

The price that one has to pay for unifying gauge theories
with gravity is the embedding of string theory in a ten
or twenty-six dimensional target-space\footnote{or some other
effective way of accounting for the same number of degrees of
freedom}. The consequence is an enormous freedom and an enormous
number of vacua when compactifying to four dimensions.
This outcome has led some authors to advocate the anthropic principle as a
possible resolution for the contrived set of parameters that seems to
govern our universe \cite{anthropics}.
A particular worry of these authors is the value of the cosmological
constant that seem to require a large degree of fine tuning.
But the primary source of confusion is the apparent multitude
of string vacua, and the lack of a mechanism to choose among them.

The problem, however, may lie elsewhere.
The basic misconception is on what information we may reliably 
extract from string theory in its contemporary level of understanding.
String theory at present is ill suited to address
the issue of the cosmological constant. What is at stake is the
basic formulation of a quantum space--time theory in a non--fixed background.
Our present understanding of string theory does not provide
that, and therefore trying to use it to address issues that are
intrinsically dynamical in their nature is at present not a
well posed problem. Nevertheless, the theory is useful
in probing the properties of quantum gravity in its static limits.
As such one can extract useful information from it
and try to isolate some basic features of the underlying theory.
This program has been extremely successful in several respects.
On the one hand we do have a viable framework in which we can
calculate the spectrum reliably and confront with the observed
spectrum, leading to models that look tantalizingly close to the
real deal. On the other hand, over the years substantial progress
has been made on the understanding of the underlying theory.
And finally, there does not exist a contemporary competitive
that can be taken seriously as achieving an equal measure of
maturity and success in providing a framework for a unifying theory.  
However, the ultimate formulation of string theory in a non--fixed
background may take a long while. 

In this article I will argue that the evidence does point in a direction
of a criteria that may be associated with the string vacuum, 
which is self--duality under the so--called T--duality transformations.
More generally I will argue that the evidence suggests that there is
an association between the vacuum state, or the classically trivial state,
and the self--dual point under phase--space duality.
The evidence arises from two completely unrelated aspects.
The first is from the fact that the most realistic string 
models to date are indeed found in the vicinity of the self--dual
point under T--duality. Thus, these phenomenological string models
motivate the hypothesis that the self--duality criteria plays
a role in the vacuum selection principle. The second is in the
framework of the equivalence postulate approach to quantum
mechanics, where one observes an association between
the trivial state and the self--dual state
under phase--space duality. These disparate consideration
may be a reflection of a very basic and general mathematical
criteria for physical systems that exhibit a duality structure,
which associates critical points in phase--space with
the self--dual points under the associated duality transformations.

\section{Some basic properties of string theory}\label{post}

String theory is a mundane modification of the concept of a relativistic
point particle. It assumes that in addition to proper time,
which parameterizes the world--line of a relativistic point particle,
one has to incorporate a parameter for an intrinsic internal dimension.
Thus, rather than world--line, the particle motion is characterized
by a world--sheet. The beauty in this simple modification
is reflected in both the elegance and power of string theory.
It motivates the hypothesis that the core of the theory
is relevant for the unified description of the fundamental
matter and interactions. 

The classical string theory exhibits the basic property that it is 
always possible to gauge fix the two dimensional world--sheet metric
to the flat metric \cite{reviewsofstringtheory}. This basic property
is unique to a one--dimensional extended object, {\it i.e.} a string.
Requiring that this fundamental property of the string is also maintained
in the quantized theory imposes a strong constraint on the
theory, which results in important phenomenological consequences.
It necessitates the embedding of the bosonic string in twenty-six
dimensions and of the fermionic string in 10 dimensions. These 
extra degrees of freedom enable the embedding of the Standard Model spectrum
into string theory. Whatever
shape the extra dimensions will take in the final theory,
the predictive power of the theory should be noted. It
predicts an exact number for the extra degrees of freedom
needed to maintain the world--sheet gauge fixing property.
Additionally it leads to the requirement of modular invariance,
which imposes further constraints on the quantized theory.

The equation of motion of the string is a two dimensional
wave equation, whose solutions are left-- and right--moving modes,
that also depend on the particular boundary conditions imposed.
In the case of the closed string the left-- and right--moving modes
are decoupled and, up to some consistency constraints, one can assign
independent boundary conditions to the left-- and right--moving modes.
The decoupling of the left and right--moving modes allows also
the possibility, which is realized in the heterotic--string
\cite{heterotic}, of having a supersymmetric left--moving sector,
whereas the right--moving sector is nonsusymmetric.
In this case, sixteen of the extra dimensions of the
nonsupersymmetric sector are compactified on a sixteen-dimensional
even self--dual lattice, whereas six right--moving coordinates
in combination with six left--moving coordinates are 
compactified on a six--dimensional complex or real 
manifold. Alternatively, one can formulate the compactified
theory directly in four dimensions by identifying the 
compactified dimensions as internal conformal field theories
propagating on the string world--sheet, and subjected 
to the string consistency requirements.
At the end of the day we expect the different formulations
to be related, and they provide different sets of tools to
study the properties of string compactifications, but do not
represent distinct physical objects.

It is noted that classical geometry, {\it e.g.} which
is realized on conventional Calabi--Yau manifolds, necessitates
that the assignment of boundary conditions to the compactified
dimensions is symmetric between the left and the right--movers.
However, the possibility
of having asymmetric boundary conditions in string theory
may play a pivotal role in the moduli fixing and vacuum selection
mechanism. 

\subsection{T--duality}

String theory exhibits various forms of dualities, {\it i.e.} 
relation between different theories at large and small radii
of the compactified manifold and at strong and weak coupling. 
The first type is the T--duality \cite{tduality}. Consider a point particle 
moving on a compactified dimension $X$, which obeys the condition
$X\sim X +2\pi R m$. Single valuedness of the wave function of the 
point particle $\Psi\sim {\rm Exp}(iP X )$ implies that the momenta
in the compact direction is quantized $P={m\over R}$ with $m\in Z$. 
Now consider a string moving in the compactified direction.
In this case the string can wrap around the compactified dimension
and produce stable winding modes. Hence the left and right--moving
momenta in the case of the closed string are given by
$$P_{L,R}={1\over{\sqrt2}}({m\over R}\pm {{nR}\over \alpha^\prime})$$
and the mass of the string states is given by
$${\rm mass}^2=\left({m\over R}\right)^2+\left({{n R}\over
\alpha^\prime}\right)^2$$
this is invariant under exchange of large and small radius together 
with the exchange of winding and momentum modes, {\it i.e.}
$${1\over R}\leftrightarrow {R\over\alpha^\prime}~~~{\rm with}~~~
m\leftrightarrow n$$
and is an exact symmetry in string perturbation theory. Furthermore, 
there exist the self--dual point, 
$$ R={\alpha^\prime\over R},$$
which is the symmetry point 
under T--duality.


The perturbative T--duality symmetry is a characteristic property
of the compactified string. The existence of a symmetry point
under this duality, namely the self--dual point, suggests that
this point may play a role in the vacuum selection. Naturally,
we would like to know what is
the dynamical mechanism that selects the vacuum. But the first
step is to try to gain further understanding of the basic
properties of such special points in the moduli space. 
An important consequence is the
emergence of winding modes that become massless at the
self--dual radius and hence enhance the symmetry. For a single
compactified coordinate the symmetry is enhanced from $U(1)$
to $SU(2)$. 

It is well known that for specific
values of its  radius, the compactified coordinate can be realized 
as specific rational conformal field theories propagating on
the string world--sheet. In particular, there exist such a value for
which a compactified coordinate can be represented in terms
of two free Majorana--Weyl fermions. It so happens that, in some
normalization, the self--dual point is at $R=1/\sqrt2$ whereas the
free fermionic point is at $R=1$. Hence, the two points do not overlap
and the free fermionic point does not coincide with the self--dual point
\cite{ginsparg}.

However, this is merely an artifact of the fact that we have been talking
here about a single compactified bosonic coordinate, which corresponds 
to two real free fermions on the world--sheet.
In this case the fermionic excitations cannot enhance the symmetry and
the two points therefore do not coincide. In higher dimensions
the situation is more intricate, and some caution should be exercised.
With two compactified coordinates the symmetry is enhanced from
$U(1)^2$ to $SO(4)\sim SU(2)\times SU(2)$, which is precisely the
symmetry, which is realized at the self--dual point, and hence the
free fermionic realization coincides with the compactified dimensions
at the self--dual point. 

This situation merits further investigation. If we proceed to six
compactified dimensions the free fermionic realization gives rise
to the maximal $SO(12)$ enhanced symmetry, whereas naively we
would expect that fixing the compactified radii at the self--dual
point enhances the symmetry to $SU(2)^6$. However,
the situation is more subtle. In higher dimensions in addition
to the compact space metric the string action allows for a
non--vanishing antisymmetric tensor field. The action for the
D--dimensional compactified string is given by,
$$
S={1\over{8\pi}}
\int{d^2\sigma({G_{ij}\partial^\alpha{X^i}\partial_\alpha{X^j}+
\epsilon^{\alpha\beta}B_{ij}\partial_\alpha{X^i}\partial_\beta{X^j}})}~,$$
where,
$$G_{ij}={1\over2}{\sum_{I=1}^D}R_ie_i^IR_je_j^I~,$$
is the metric of the six dimensional compactified space 
and $B_{ij}=-B_{ji}$ is the antisymmetric tensor field. 
The $e^i=\{e_i^I\}$ are six linear independent vectors normalized
to $(e_i)^2=2$.
The left-- and right--moving momenta are given by,
\beq
P^I_{R,L}=[m_i-{1\over2}(B_{ij}{\pm}G_{ij})n_j]{e_i^I}^*~,
\label{lrmomenta}
\eeq
where the ${e_i^I}^*$ are dual to the $e_i$, and  
$e_i^*\cdot e_j=\delta_{ij}$. The left-- and right--moving momenta span a 
Lorentzian even self--dual lattice. The mass formula for the left and 
right--movers is,
$$M_L^2=-c+{{P_L\cdot{P_L}}\over2}+N_L=-1+{{P_R\cdot{P_R}}\over2}+
N_R=M_R^2~,$$
where $N_{L,R}$ are the sum on the left--moving and right--moving oscillators
and $c$ is a normal ordering constant equal to ${1\over2}$ and $0$
for the antiperiodic (NS) and periodic (R) sectors of the NSR fermions. 

The T--duality symmetry of string theory compactified on a
D--dimensional manifold is enlarged to \cite{tduality},
$$G+B~\rightarrow~{1\over{G+B}},$$
where $G$ and $B$ are the metric and the antisymmetric tensor
of the $D$--dimensional compactified manifold respectively. 
We can then distinguish between the self--dual points
in the moduli space and the points of maximally enhanced symmetry.
For a six dimensional compactified space
the maximally symmetric point is the $SO(12)$ lattice,
which is produced at the free fermionic point.
Of further interest is the relation between
the points of maximally enhanced symmetry and the self--dual
points. ``maximal'' here refers to an enhanced semi--simple
and simply--laced symmetry group of rank $D$ corresponding
to a level 1 affine Lie algebra with maximal dimensionally.
In practice these are $SU(D+1)$ or $SO(2D)$ gauge groups,
as spinorial representations under these groups are not
realized in the internal manifold, and hence enhancement
to an exceptional group does not occur. In table \ref{denhancement}
the dimensionality of the possible enhanced symmetries
up to $D=6$ is listed. 
\begin{eqnarray}
 &\begin{tabular}{|c|c|c|}
\hline
$D$ & $SU(D+1)$ & $SO(2D)$ \\
\hline
 1  &    3      &     1    \\
 2  &    8      &     6    \\
 3  &    15     &     15   \\
 4  &    24     &     28   \\
 5  &    35     &     45   \\
 6  &    48     &     66   \\
\hline
\end{tabular}
\label{denhancement}
\end{eqnarray}
{}At the free fermionic point with $D\ge 2$ the resulting enhanced symmetry
is $SO(2D)$. From table \ref{denhancement} we note that this is the 
maximally enhanced symmetry for $D\ge3$, whereas for $D=2$ the free
fermionic point coincides with the self--dual point, and is not at the
point of maximally enhanced symmetry.

As already noted above for $D=1$ moduli space the self--dual point of
circle compactification coincides with the appearance of the enhanced
$SU(2)$ gauge symmetry\footnote{in the heterotic string the enhanced
gauge symmetry is realized only on the non--supersymmetric side.}.
Denoting $E=G+B$, the duality transformation $E\rightarrow E^{-1}$
still has strictly speaking a single self--dual point given by
$$G~=~I~,~~~~B~=~0,$$
which is the unique solution of the equation $(G+B)^2=I$
when $G$ is positive definite. At this self--dual point the
string forms a level--one representation of the affine
algebra $SU(2)^D$. Considering the more general case of
self--dual points of the transformations $E~\rightarrow~ E^{-1}$,
modulo $SL(D,Z)$ and $\Theta(Z)$ transformations \cite{grv,tduality}
$$E^{-1}~=~M^t(E+\Theta)M~~,~~M\in SL(D,Z)~,~\Theta\in\Theta(Z)~.$$
In ref. \cite{grv,tduality} it is demonstrated that any background
with maximally enhanced symmetry falls into this category.
In those cases the background is \cite{egrs},
$$E_{ij}=2 C_{ij} ~~i~>~j~~,~~
  E_{ii}= C_{ii} ~~i~>~j~~,~~
  E_{ij}=o ~~i~<~j~~,~~$$
where $C_{ij}$ is the Cartan matrix. Therefore, 
$$E,E^{-1}\in SL(D,Z),$$
and $M=E^{-1},~\Theta=E^t-E$. This shows that a maximally enhanced
symmetry point is a self--dual point under some non--trivial
$O(D,D,Z)$ transformation, and therefore an orbifold point in the
moduli space.

The next point in the discussion is therefore to find
the orbifold transformations that reproduce the
$SO(12)$ lattice, which is generated at the free fermionic
point for $D=6$. The background fields that produce
the toroidal $SO(12)$ lattice are given by the metric,
\beq
g_{ij}=\left(\matrix{~2&-1& ~0& ~0& ~0& ~0\cr
-1& ~2&-1& ~0& ~0& ~0\cr~0&-1& ~2&-1& ~0& ~0\cr~0& ~0&-1
& ~2&-1&-1\cr ~0& ~0& ~0&-1& ~2& ~0\cr ~0& ~0& ~0&-1& ~0& ~2\cr}\right)~,
\label{gso12}
\eeq
and the antisymmetric tensor,
\beq
b_{ij}=\cases{
g_{ij}&;\ $i>j$,\cr
0&;\ $i=j$,\cr
-g_{ij}&;\ $i<j$.\cr}
\label{bso12}
\eeq
When all the radii of the six-dimensional compactified
manifold are fixed at $R_I=\sqrt2$, it is seen that the
right--moving momenta given by eqs. (\ref{lrmomenta})
produce the root vectors of $SO(12)$. 

The realization of the $SO(12)$ lattice as an orbifold
in achieved by incorporating idenitifications on the 
internal lattice by shift symmetries. It is instructive
for this purpose to study the partition function at a 
generic point in the moduli space, incorporate the 
shifts, and fix the internal radii at the self--dual
point, which then reproduces the partition function
of the $SO(12)$ lattice. The partition function at the
of the $N=4$ supersymmetric $SO(12)\times E_8\times E_8$
heterotic vacuum is given by
\beq
{\Scr Z}=(V_8-S_8)\left[|O_{12}|^2+|V_{12}|^2+|S_{12}|^2+|C_{12}|^2\right]
\left( \bar O_{16} + \bar S_{16}\right) \left( \bar O_{16} + \bar S_{16}
\right) \,, \label{zplus}
\eeq
where ${\Scr Z}$ has been written in terms of level-one
${\rm SO} (2n)$ characters (see, for instance, \cite{as})
\beqn
O_{2n} &=& {\textstyle{1\over 2}} \left( {\vartheta_3^n \over \eta^n} +
{\vartheta_4^n \over \eta^n}\right) \,,
\nonumber \\
V_{2n} &=& {\textstyle{1\over 2}} \left( {\vartheta_3^n \over \eta^n} -
{\vartheta_4^n \over \eta^n}\right) \,,
\nonumber \\
S_{2n} &=& {\textstyle{1\over 2}} \left( {\vartheta_2^n \over \eta^n} +
i^{-n} {\vartheta_1^n \over \eta^n} \right) \,,
\nonumber \\
C_{2n} &=& {\textstyle{1\over 2}} \left( {\vartheta_2^n \over \eta^n} -
i^{-n} {\vartheta_1^n \over \eta^n} \right) \,.
\eeqn

On the compact coordinates there are actually three inequivalent ways
in which the shifts
can act. In the more familiar case, they simply translate a generic point 
by half the
length of the circle. As usual, the presence of windings in string 
theory allows shifts on the T-dual circle, or even asymmetric ones, that 
act both on the circle and on its dual. More concretely, for a circle of
length $2 \pi R$, one can have the following options \cite{vwaaf}:
\beqn
A_1\;:&& X_{\rm L,R} \to X_{\rm L,R} + {\textstyle{1\over 2}} \pi R \,,
\nonumber \\
A_2\;:&& X_{\rm L,R} \to X_{\rm L,R} + {\textstyle{1\over 2}} \left(
\pi R \pm {\pi \alpha ' \over R} \right) \,, 
\nonumber \\
A_3\;:&& X_{\rm L,R} \to X_{\rm L,R} \pm {\textstyle{1\over 2}} {\pi \alpha'
\over R} \,.
\label{a1a2a3}
\eeqn
There is, however, a crucial difference among these three choices: while
$A_1$ and $A_3$ shifts can act consistently on any number of coordinates,
level-matching requires instead that the $A_2$-shifts act on (mod) four real 
coordinates. 

Our problem is to find the shift that when acting on the 
lattice $T_2^1\otimes T_2^2\otimes T_2^3$ at a generic point in
the moduli space reproduces the $SO(12)$ lattice when the radii
are fixed at the self--dual point $R=\sqrt{\alpha^\prime}$ \cite{partitions}.
Let us consider for simplicity the case of six orthogonal circles or 
radii $R_i$. The partition function reads
\beq
{\Scr Z}_+ = (V_8 - S_8) \, \left( \sum_{m,n} \Lambda_{m,n}
\right)^{\otimes 6}\, \left( \bar O _{16} + \bar S_{16} \right) \left(
\bar O _{16} + \bar S_{16} \right)\,,
\eeq
where as usual, for each circle,
\beq
p_{\rm L,R}^i = {m_i \over R_i} \pm {n_i R_i \over \alpha '} \,,
\eeq
and
\beq
\Lambda_{m,n} = {q^{{\alpha ' \over 4} 
p_{\rm L}^2} \, \bar q ^{{\alpha ' \over 4} p_{\rm R}^2} \over |\eta|^2}\,.
\eeq
We can now act with the $\bb{Z}_2\times \bb{Z}_2$ shifts generated by
\beqn
g\;: & & (A_2 , A_2 ,0 ) \,,
\nonumber \\
h\;: & & (0, A_2 , A_2 ) \,, \label{gfh}
\eeqn
where each $A_2$ acts on a complex coordinate. The resulting partition 
function then reads
\beqn
{\Scr Z}_+ &=& {\textstyle{1\over 4}}\, (V_8 - S_8) 
\sum_{m_i , n_i}  \left\{ \left[ 1 + (-1)^{m_1 + m_2 + m_3 + m_4 + n_1 + n_2 +
n_3 + n_4} \right. \right.
\nonumber \\
& & \left. + (-1)^{m_1 + m_2 + m_5 + m_6 + n_1 + n_2 +
n_5 + n_6} + (-1)^{m_3 + m_4 + m_5 + m_6 + n_3 + n_4 +
n_5 + n_6} \right]  
\nonumber \\
& & \left. \times \left( \Lambda_{m_i , n_i}^{1,\ldots ,6} 
+ \Lambda^{1,\ldots,4}_{m_i + {1\over 2}, n_i + {1\over 2}} 
\Lambda^{5,6}_{m_i , n_i} 
+ \Lambda^{1,2,5,6}_{m_i + {1\over 2}, n_i + {1\over 2}} 
\Lambda^{3,4}_{m_i , n_i}
+ \Lambda^{1,2}_{m_i , n_i} 
\Lambda^{3,4,5,6}_{m_i + {1\over 2}, n_i + {1\over 2}} 
\right) \right\}
\nonumber \\
& & \times
\left( \bar O _{16} + \bar S_{16} \right) \left( \bar O_{16} + \bar S_{16}
\right) \label{zpshift}
\eeqn

After some tedious algebra, it is then possible to show that, once evaluated
at the self-dual radius $R_i = \sqrt{\alpha '}$, the 
partition function (\ref{zpshift}) reproduces that at the SO(12) point
(\ref{zplus}). To this end, it suffices to notice that
\beqn
\sum_{m,n} \Lambda_{m,n} (R=\sqrt{\alpha '}) &=& |\chi_0 |^2 + 
|\chi_{1\over 2}
|^2 \,,
\nonumber \\
\sum_{m,n} (-1)^{m+n} \Lambda_{m,n} (R = \sqrt{\alpha '}) &=&
|\chi_0 |^2 - |\chi_{1\over 2} |^2 \,,
\nonumber \\
\sum_{m,n} \Lambda_{m + {1\over 2} , n + {1\over 2}} (R = \sqrt{\alpha '})
&=& \chi_0 \bar\chi_{1\over 2} + \chi_{1\over 2} \bar \chi_0 \,,
\nonumber \\
\sum_{m,n} (-1)^{m+n} \Lambda_{m + {1\over 2} , n + {1\over 2}} 
(R = \sqrt{\alpha '}) &=& \chi_{1\over 2} \bar \chi_0 -
\chi_0 \bar\chi_{1\over 2}  \,,
\eeqn
where
\beqn
\chi_0 &=& \sum_\ell q^{\ell^2} \,,
\nonumber \\
\chi_{1\over 2} &=& \sum_\ell q^{(\ell + {1\over 2})^2} \,,
\eeqn
are the two level-one SU(2) characters, while, standard branching rules,
decompose the SO(12) characters into products of SU(2) ones. For instance,
\beqn
O_{12} &=& \chi_0 \chi_0 \chi_0 \chi_0 \chi_0 \chi_0 +
\chi_0 \chi_0 \chi_{1\over 2} \chi_{1\over 2}
\chi_{1\over 2} \chi_{1\over 2}
\nonumber \\
& &\chi_{1\over 2} \chi_{1\over 2} \chi_0 
\chi_0 \chi_{1\over 2} \chi_{1\over 2} +
\chi_{1\over 2} \chi_{1\over 2} \chi_{1\over 2} \chi_{1\over 2} 
\chi_0 \chi_0 \,.
\eeqn

To summarize this section, string theory is a rather mundane
modification of the concept of a relativistic point particle.
Remarkably, consistency of the quantized string
necessitates that the string is embedded in higher dimensions,
some of which must be compactified to conform with reality.
Propogation of the perturbative string in the compact directions
exhibits invariance under T--duality transformations, as well
as enhanced symmetries for critical values of the compact
coordinates. The points of maximally enhanced symmetries
coincide with self--dual points under the T--duality transformations
up to $SL(D,Z)$ and $\Theta(Z)$ transformations.
The natural expectation is that the self--dual point
in the moduli space will have a physical significance.
In the following I turn to a class of orbifolds that
are contructed in the vicinity of the self--dual point.

\section{Realistic string models}\label{rsm}

{}From the Standard Model data we may hypothesis that the string vacuum
should possess two key properties. The existence of three generations and
their embedding in $SO(10)$ representations. The only perturbative string
theory that preserves the $SO(10)$ embedding is the heterotic string,
because this is the only one that produces the chiral 16 representation
of $SO(10)$ in the perturbative spectrum. This is an important
phenomenological qualification of the different perturbative string theories.
In this respect, it may well be that other phenomenological
properties of the Standard Model spectrum will be more readily
accessible in other M--theory limits. 

The exploration of realistic superstring vacua proceeds by studying
compactification of the heterotic string from ten to four dimensions. 
There is a large number of possibilities. The first type of semi--realistic
superstring vacua that were constructed are compactification
on Calabi--Yau manifolds that give rise to an $E_6$ observable
gauge group, which is broken further by the
Hosotani flux breaking \cite{hosotani}
mechanism to $SU(3)^3$ \cite{suthree}. This gauge group is then broken
to the Standard Model gauge group in the effective field theory level. 
This type of geometrical compactifications correspond at special
points to conformal world--sheet field theories,
which have $N=2$ world--sheet supersymmetry in the left-- and right--moving
sectors. Similar geometrical compactifications
which have only (2,0) world--sheet supersymmetry have also been
studied and can lead to compactifications with $SO(10)$ and $SU(5)$
observable gauge group \cite{twozero}.
The analysis of this type of compactification
is complicated due to the fact that they do not correspond
to free world--sheet theories. Therefore, it is rather difficult
to calculate the spectrum and the parameters of the Standard Model
in these compactifications. On the other hand they may provide
a sophisticated mathematical window to the underlying quantum geometry. 

The next class of superstring vacua that have been explored in detail are
the orbifold models \cite{dhvw}.
In these models one starts with a compactification
of the heterotic string on a flat torus,
using the Narain prescription \cite{narain}.
This type of compactifications uses free world--sheet bosons. The Narain
lattice is then moded out by some discrete symmetries which are the
orbifold twistings. The most detailed study of this type of models are
the $Z_3$ orbifold \cite{zthree},
which give rise to three generation models with
$SU(3)\times SU(2)\times U(1)^n$ gauge group. One caveat of this class of
models is that the weak--hypercharge does not have the standard
$SO(10)$ embedding.
Thus, the nice features of $SO(10)$ unification are lost. This fact has
a crucial implication that the normalization of the weak hypercharge
relative to the non--Abelian currents is larger than 5/3, the standard
$SO(10)$ normalization. This results generically in disagreement with the
observed low energy values for $\sin^2\theta_W(M_Z)$ and $\alpha_s(M_Z)$. 

A class of heterotic string models that accommodates three generations and
the $SO(10)$ embedding of the Standard Model spectrum, are the so called
free fermionic models. As noted above the free fermionic point in the moduli
space of superstring theories is related to the self--dual point. The other
key property of the  free fermionic models is their relation to
$Z_2\times Z_2$ orbifold compactification. While the space of possible
string compactifications may be beyond count, any model, or class of models,
that exhibits realistic properties deserve to be studied in depth. 

The class of three generation free fermionic models is therefore 
constructed precisely in the vicinity of the self--dual point under 
T--duality! This is an extremely intriguing coincidence!
The structure of the $Z_2\times Z_2$ orbifold naturally
correlates the existence of three generations
with the underlying geometry. This arises due to the fact that 
the $Z_2\times Z_2$ orbifold has exactly three twisted sectors. 
Each of the light chiral generations then arises from a distinct 
twisted sector\footnote{2+1 three generation models,
with two generations arising from one twisted sector and the third
arising from another, are also possible \cite{twoplusone,nilles}}.
Hence, in these models the existence of three generations
in nature is seen to arise due to the fact that we are dividing
a six dimensional compactified manifold into factors of 2. In simplified
terms, three generations is an artifact of 
$$
{6\over 2}~~=~~1~+~1~+~1~.~
$$
One may further ask whether there is a reason that the $Z_2$ 
orbifold would be preferred versus higher orbifolds.
Previously I argued that the free fermionic point 
is identified with the self--dual point under $T$--duality,
which is where we would naively expect the compactified 
dimensions to stabilize. The special property of the 
$Z_2$ orbifold that sets it apart from higher orbifolds,
is the fact that the $Z_2$ orbifold is the only one that acts
on the coordinates as real coordinates rather than complex 
coordinates. The class of string vacua that we are led to consider
are $Z_2\times Z_2$ orbifolds in the vicinity of the self--dual
point of the six dimensional compactified space.
In the vicinity of this point, the compactified dimensions
can be represented in terms of free fermions propagating 
on the string world--sheet, and deformations from
the self--dual point correspond to the inclusion of
world--sheet Thirring interactions. 

\section{Free fermionic model building}\label{ffmb}
The three generation $Z_2\times Z_2$ orbifold models were studied
in the free fermionic formulation \cite{fff}. These models were reviewed
in \cite{btd979902}, and I give here a brief summary of their main properties.
The models are constructed in terms of a set of boundary
condition basis vectors that define the transformation properties 
of the 20 left--moving and 44 right--moving real fermions
around the noncontractible loops of the one--loop vacuum to vacuum 
amplitude. 

The first five basis vectors of the realistic free fermionic
models consist of the NAHE set \cite{nahe}, $\{1, S, b_1, b_2, b_3\}$.
The gauge group after the NAHE set is
$SO(10)\times E_8\times SO(6)^3$ with $N=1$ space--time supersymmetry,
and 48 spinorial $16$ of $SO(10)$, sixteen from each sector $b_1$,
$b_2$ and $b_3$. The three sectors $b_1$, $b_2$ and $b_3$ are
the three twisted sectors of the corresponding $Z_2\times Z_2$  
orbifold compactification. 

The NAHE set is common to a large class of three generation
free fermionic models. The construction proceeds by adding to the
NAHE set three additional boundary condition basis vectors, typically
denoted by $\{\alpha,\beta,\gamma\}$,
which break $SO(10)$ to one of its subgroups: $SU(5)\times U(1)$
\cite{revamp}, $SO(6)\times SO(4)$ \cite{patisalamstrings},
$SU(3)\times SU(2)\times U(1)^2$ \cite{fny,eu,top,cfn},
$SU(3)\times U(1)\times SO(4)$ \cite{lrsstringmodels},
or $SU(4)\times SU(2)\times U(1)$ \cite{su421}.
At the same time the number of generations is reduced to
three, one from each of the sectors $b_1$, $b_2$ and $b_3$.
The various three generation models differ in their
detailed phenomenological properties. However, many of  
their characteristics can be traced back to the underlying
NAHE set structure. One such important property to note
is the fact that as the generations are obtained  
from the three twisted sectors $b_1$, $b_2$ and $b_3$,
they automatically possess the Standard $SO(10)$ embedding.
Consequently the weak hypercharge, which arises as
the usual combination $U(1)_Y=1/2 U(1)_{B-L}+ U(1)_{T_{3_R}}$,
has the standard $SO(10)$ embedding.

The massless spectrum of the realistic free fermionic models
then generically contains three generations from the
three twisted sectors $b_1$, $b_2$ and $b_3$, which are
charged under the horizontal symmetries. The Higgs spectrum
consists of three pairs of electroweak doublets from the
Neveu--Schwarz sector plus possibly additional one or
two pairs from a combination of the two basis vectors  
which extend the NAHE set. Additionally the models
contain a number of $SO(10)$ singlets which are
charged under the horizontal symmetries and 
a number of exotic states.

Exotic states arise from the basis vectors which extend the NAHE
set and break the $SO(10)$ symmetry \cite{ccf}. Consequently, they
carry either fractional $U(1)_Y$ or $U(1)_{Z^\prime}$ charge.
Such states are generic in superstring models and impose severe constraints
on their validity. In some cases the exotic fractionally charged
states cannot decouple from the massless spectrum, and their presence
invalidates otherwise viable models \cite{otherrsm}. In the NAHE based
models the fractionally charged states always appear in vector--like
representations. Therefore, in general mass terms are generated from
renormalizable or nonrenormalizable terms in the superpotential.
However, the mass terms which arise from non--renormalizable terms will in
general be suppressed, in which case the fractionally charged states may have
intermediate scale masses. The analysis of ref. \cite{cfn} demonstrated the  
existence of free fermionic models with solely the  MSSM spectrum in the low
energy effective field theory of the Standard Model charged matter.

\section{Phenomenological studies of free fermionic models}

I summarize here some of the highlights of the phenomenological studies
of the free fermionic models. This demonstrates that the free fermionic
string models indeed provide the arena for exploring many of the questions
relevant for the phenomenology of the Standard Model and Unification.
Hence, the underlying structure of these models, generated by the NAHE set,
produces the right features for obtaining realistic phenomenology. It
provides further evidence for the assertion that the true string vacuum is
connected to the $Z_2\times Z_2$ orbifold in the vicinity of the self--dual
point in the Narain moduli space. Many of the important issues relating to
the phenomenology of the Standard Model and supersymmetric unification have
been discussed in the past in several prototype free fermionic
heterotic string models. These studies are reviewed in
\cite{btd979902}, where further details can be found.
These include among others: top quark mass prediction \cite{top}, several
years prior to the actual observation by the CDF/D0 collaborations
\cite{cdfd0};
generations mass hierarchy \cite{NRT}; CKM mixing \cite{CKM};
superstring see--saw mechanism \cite{seesaw}; Gauge coupling
unification \cite{gcu}; Proton stability \cite{ps};
supersymmetry breaking and squark degeneracy \cite{fp2,dedes}.
Additionally,
it was demonstrated in ref. \cite{cfn} that at low energies the model
of ref. \cite{fny}, which may be viewed as a
prototype example of a realistic free fermionic model,
produces in the observable sector solely the MSSM charged spectrum.
Therefore, the model of ref. \cite{fny}, supplemented
with the flat F and D solutions of ref. \cite{cfn}, provided
the first example in the literature of a string model
with solely the MSSM charged spectrum
below the string scale. Thus, for the first time it provided
an example of a long--sought Minimal Superstring Standard Model!
We have therefore identified
a neighborhood in string moduli space which is potentially
relevant for low energy phenomenology. While we can suggest 
arguments, based on target--space duality considerations why this
neighborhood may be selected, we cannot credibly argue that
similar results cannot be obtained in other regions of the string moduli
space. Nevertheless, these results provide the justification for further
explorations of the free fermionic models.
In this respect, the vital property of the free fermionic models
is their connection with the $Z_2\times Z_2$ orbifold, to which I turn in
section \ref{z2z2orbifold}.

I would like to emphasize that it is not suggested that any of the
realistic free fermionic models is the true vacuum of our world.
Indeed such a claim would be folly. Each of the phenomenological
free fermionic models has its shortcomings. In particular, their
does not exist a demonstration of a single model
that incorporates all of
the phenomenological constraints imposed by the Standard Model
data with a single choice of flat directions. 
While in principle the phenomenology of each of these models may be improved
by further detailed analysis of supersymmetric flat directions,
it is not necessarily the most interesting avenue for exploration.
The aim of the studies outlined above is to demonstrate that
all of the major issues, pertaining to the phenomenology of the
Standard Model and unification, can in principle be
addressed in the framework of the free fermionic models,
rather than to find the explicit solution that accommodates
all of these requirements simultaneously. The reason being
that even within this space of solutions there is still a vast
number of possibilities, and we still lack the guide to select the
most promising one. The question which is of interest is whether
there are some deeper reasons that would indicate why the 
free fermionic models may be preferred. The argument of this
paper is that self--duality under T--duality, or more generally
self--duality in quantum phase space, is the fundamental principle
that is associated with the quantum gravity vacuum selection
mechanism. Thus, the phenomenological guide provided by the
free fermionic models, may lead to deeper insight into the
basic properties of string theory and quantum gravity. 
This perspective provides the motivation for the continued interest in
the detailed study of this class of string compactifications.

The free fermionic models also serve as a laboratory to study 
possible signatures beyond the Standard Model. 
These include the possibility of extended gauge symmetries \cite{zp};
specific supersymmetric spectrum scenarios \cite{dedes}; 
and exotic matter \cite{ccf}. 
Perhaps most fascinating among
those is the existence of exotic matter states \cite{fccp,ccf}
that can lead to experimental signatures in the form of
energetic neutrinos from the sun \cite{fop}, or in the form
of candidates for dark matter and top--down UHECR scenarios \cite{cfp}.
The later is particularly exciting due to the forthcoming Pierre Auger and
EUSO experiments that will provide more statistics on UHECR. 

\section{Correspondence with $Z_2\times Z_2$ orbifold}\label{z2z2orbifold}

The key property of the realistic free fermionic models is
the correspondence with the $Z_2\times Z_2$ orbifold compactification. 
As discussed in section \ref{ffmb} the construction of the
realistic free fermionic models can be divided into two
parts. The first part consist of the NAHE--set basis vectors, 
$\{1, S, b_1, b_2, b_3\}$,
and the second consists of the additional boundary conditions,
$\{\alpha,\beta,\gamma\}$.
The correspondence of the NAHE-based free fermionic models  
with the orbifold construction is illustrated
by extending the NAHE set, $\{ 1,S,b_1,b_2,b_3\}$, by one additional   
boundary condition basis vector \cite{foc},
\beq
\xi_1=(0,\cdots,0\vert{\underbrace{1,\cdots,1}_{{\bar\psi^{1,\cdots,5}},
{\bar\eta^{1,2,3}}}},0,\cdots,0)~.
\label{vectorx}
\eeq
With a suitable choice of the GSO projection coefficients the
model possesses an ${\rm SO}(4)^3\times {\rm E}_6\times {\rm U}(1)^2
\times {\rm E}_8$ gauge group
and $N=1$ space-time supersymmetry. The matter fields
include 24 generations in the 27 representation of
${\rm E}_6$, eight from each of the sectors $b_1\oplus b_1+\xi_1$,
$b_2\oplus b_2+\xi_1$ and $b_3\oplus b_3+\xi_1$.
Three additional 27 and $\overline{27}$ pairs are obtained
from the Neveu-Schwarz $\oplus~\xi_1$ sector.

To construct the model in the orbifold formulation one starts
with the compactification on a torus with nontrivial background
fields \cite{narain}.
The subset of basis vectors,
\beq
\{ 1,S,\xi_1,\xi_2\},
\label{neq4set}
\eeq
generates a toroidally-compactified model with $N=4$ space-time
supersymmetry and ${\rm SO}(12)\times {\rm E}_8\times {\rm E}_8$ gauge
group.
The construction of this $N=4$ string vacuum in the geometric (bosonic)
language was discussed in section \ref{post}.

Adding the two basis vectors $b_1$ and $b_2$ to the set
(\ref{neq4set}) corresponds to the ${Z}_2\times {Z}_2$
orbifold model with standard embedding.
Starting from the Narain model with ${\rm SO}(12)\times
{\rm E}_8\times {\rm E}_8$
symmetry~\cite{narain}, and applying the ${Z}_2\times {Z}_2$
twist on the
internal coordinates, reproduces
the spectrum of the free-fermion model
with the six-dimensional basis set
$\{ 1,S,\xi_1,\xi_2,b_1,b_2\}$.
The Euler characteristic of this model is 48 with $h_{11}=27$ and
$h_{21}=3$. I denote the manifold corresponding to this
model as $X_2$.

It is noted that the effect of the additional basis vector $\xi_1$ of eq.
(\ref{vectorx}), is to separate the gauge degrees of freedom, spanned by
the world-sheet fermions $\{{\bar\psi}^{1,\cdots,5},
{\bar\eta}^{1},{\bar\eta}^{2},{\bar\eta}^{3},{\bar\phi}^{1,\cdots,8}\}$,
from the internal compactified degrees of freedom $\{y,\omega\vert
{\bar y},{\bar\omega}\}^{1,\cdots,6}$.
In the realistic free fermionic
models this is achieved by the vector $2\gamma$ \cite{foc}, with
\beq
2\gamma=(0,\cdots,0\vert{\underbrace{1,\cdots,1}_{{\bar\psi^{1,\cdots,5}},
{\bar\eta^{1,2,3}} {\bar\phi}^{1,\cdots,4}} },0,\cdots,0)~,
\label{vector2gamma}
\eeq
which breaks the ${\rm E}_8\times {\rm E}_8$ symmetry to ${\rm
SO}(16)\times
{\rm SO}(16)$.
The ${Z}_2\times {Z}_2$ twist breaks the gauge symmetry to
${\rm SO}(4)^3\times {\rm SO}(10)\times {\rm U}(1)^3\times {\rm SO}(16)$.
The orbifold still yields a model with 24 generations,
eight from each twisted sector,
but now the generations are in the chiral 16 representation
of SO(10), rather than in the 27 of ${\rm E}_6$. The same model can
be realized with the set
$\{ 1,S,\xi_1,\xi_2,b_1,b_2\}$,
by projecting out the $16\oplus{\overline{16}}$
from the $\xi_1$-sector taking
\beq
c{\xi_1\choose \xi_2}\rightarrow -c{\xi_1\choose \xi_2}.
\label{changec}
\eeq
This choice also projects out the massless vector bosons in the
128 of SO(16) in the hidden-sector ${\rm E}_8$ gauge group, thereby
breaking the ${\rm E}_6\times {\rm E}_8$ symmetry to
${\rm SO}(10)\times {\rm U}(1)\times {\rm SO}(16)$.
The freedom in ({\ref{changec}) corresponds to
a discrete torsion in the toroidal orbifold model.
At the level of the $N=4$
Narain model generated by the set (\ref{neq4set}),
we can define two models, ${Z}_+$ and ${Z}_-$, depending on the
sign
of the discrete torsion in eq. (\ref{changec}). The first, say ${Z}_+$,
produces the ${\rm E}_8\times {\rm E}_8$ model, whereas the second, say
${Z}_-$, produces the ${\rm SO}(16)\times {\rm SO}(16)$ model.
The ${Z}_2\times {Z}_2$
twist acts identically in the two models, and their physical
characteristics
differ only due to the discrete torsion eq. (\ref{changec}).

This analysis confirms that the ${Z}_2\times {Z}_2$ orbifold on the
SO(12) Narain lattice is indeed at the core of the
realistic free fermionic models. However, it
differs from the ${Z}_2\times {Z}_2$ orbifold on
$T_2^1\times T_2^2\times T_2^3$, which gives $(h_{11},h_{21})=(51,3)$.
I will denote the manifold of this model as $X_1$.
In \cite{befnq} it was shown that the two models may be connected
by adding a freely acting twist or shift.
Let us first start with the compactified
$T^1_2\times T^2_2\times T^3_2$ torus parameterized by  
three complex coordinates $z_1$, $z_2$ and $z_3$,
with the identification
\beq
z_i=z_i + 1\,, \qquad z_i=z_i+\tau_i \,,
\label{t2cube}
\eeq
where $\tau$ is the complex parameter of each
$T_2$ torus.
With the identification $z_i\rightarrow-z_i$, a single torus
has four fixed points at
\beq
z_i=\{0,{\textstyle{1\over 2}},{\textstyle{1\over 2}}\,\tau,
{\textstyle{1\over 2}} (1+\tau) \}.
\label{fixedtau}
\eeq
With the two ${Z}_2$ twists
\beqn
&& \alpha:(z_1,z_2,z_3)\rightarrow(-z_1,-z_2,~~z_3) \,,
\cr
&&  \beta:(z_1,z_2,z_3)\rightarrow(~~z_1,-z_2,-z_3)\,,
\label{alphabeta}
\eeqn
there are three twisted sectors in this model, $\alpha$,
$\beta$ and $\alpha\beta=\alpha\cdot\beta$, each producing
16 fixed tori, for a total of 48. Adding
to the model generated by the ${Z}_2\times {Z}_2$
twist in (\ref{alphabeta}), the additional shift
\beq
\gamma:(z_1,z_2,z_3)\rightarrow(z_1+{\textstyle{1\over2}},z_2+
{\textstyle{1\over2}},z_3+{\textstyle{1\over2}})
\label{gammashift}
\eeq
produces again fixed tori from the three
twisted sectors $\alpha$, $\beta$ and $\alpha\beta$.
The product of the $\gamma$ shift in (\ref{gammashift})
with any of the twisted sectors does not produce any additional
fixed tori. Therefore, this shift acts freely.
Under the action of the $\gamma$-shift,
the fixed tori from each twisted sector are paired.
Therefore, $\gamma$ reduces
the total number of fixed tori from the twisted sectors   
by a factor of ${2}$,
yielding $(h_{11},h_{21})=(27,3)$. This model therefore
reproduces the data of the ${Z}_2\times {Z}_2$ orbifold
at the free-fermion point in the Narain moduli space.

A comment is in order here in regard to the matching of the 
model that include the shift and the model on the $SO(12)$ lattice.
We noted above that the freely
acting shift (\ref{gammashift}), added to the ${Z}_2\times 
{Z}_2$ orbifold
at a generic point of $T_2^1\times T_2^2\times T_2^3$,
reproduces the data of the ${Z}_2\times {Z}_2$
orbifold acting on the SO(12) lattice.  
This observation 
does not prove, however, that the vacuum which includes the shift
is identical to the free fermionic model. While the 
massless spectrum of the two models may coincide
their massive excitations, in general, may differ.
The matching of the massive spectra is examined by
constructing the partition function of the ${Z}_2\times {Z}_2$
orbifold of an SO(12) lattice, and subsequently
that of the model at a generic point including the
shift. In effect since the action of the ${Z}_2\times {Z}_2$
orbifold in the two cases is identical the problem
reduces to proving the existence of a freely
acting shift that reproduces the partition function of the
SO(12) lattice
at the free fermionic point. Then since the action of 
the shift and the orbifold projections are commuting
it follows that the two ${Z}_2\times {Z}_2$ orbifolds
are identical. The precise form of the orbifold shifts that
produces the $SO(12)$ lattice was discussed in section \ref{post}
and given in eq. (\ref{gfh}).  
On the other hand, the shifts given in Eq. (\ref{gammashift}),
and similarly the analogous freely acting shift given by 
$(A_3,A_3,A_3)$, do not reproduce the partition function
of the $SO(12)$ lattice. 
Therefore, the shift in eq. (\ref{gammashift}) does reproduce
the same massless spectrum and symmetries of the ${Z}_2\times {Z}_2$
at the free fermionic point, but the partition functions of the 
two models differ!

The lesson to extract from this analysis is that the chiral spectrum 
of the $Z_2\times Z_2$ orbifold can be reduced by acting with shift
identifications on the internal $T_2$ tori. Some of these shifts
may be freely acting and some may not. In this respect, it is 
instructive to classify all such possible shifts and the interesting
question is whether it is possible to reduce the number of generations
to three generations in this manner. The second interesting observation
is that, due to the commutativity of the $Z_2\times Z_2$ orbifold
twistings with the shift identifications, in a sense the chiral
content is already predetermined at the level of the $N=4$ lattice.
Here it is observed that the $Z_2\times Z_2$ orbifold on $SO(4)^3$
lattice produces 48 chiral generations, whereas its action on the
$SO(12)$ lattice produces 24 chiral generations.
This point is worthy of deeper investigation.

\section{The role of the additional basis vectors}

The free fermionic models correspond to $Z_2\times Z_2$
orbifold at an enhanced symmetry point in the Narain moduli space.
As argued above the $Z_2\times Z_2$ orbifold, via its free fermion
realization, naturally produces three generation models arising from the
three twisted sectors. However, the geometrical correspondence
of the free fermionic models is so far understood for the extended
NAHE set models, {\it i.e.} for the case of the $X_2$ manifold 
with 24 generations. Hence, in order to promote the geometrical 
understanding of the origin of the three generations in the free 
fermionic models, it is important to understand the geometrical 
interpretation of the boundary condition basis vectors beyond the 
NAHE set. 

Let us review for this purpose the vacuum structure in the twisted
sectors $b_1$, $b_2$ and $b_3$. In the light--cone gauge
the world--sheet free fermion field
content includes: in the left--moving sector the
two space--time fermions $\psi^\mu_{1,2}$ and the six real triples
$\{\chi_i, y_i,\omega_i\}$ $(i=1,\cdots,6)$; in the right--moving sector
the six real doubles $\{{\bar y}_i,{\bar\omega}_i\}$ $(i=1,\cdots,6)$
and the sixteen complex fermions
$\{{\bar\Psi}^{1,\cdots,5},{\bar\eta}^{1,2,3},{\bar\phi}^{1,\cdots,8}\}$. 
For our purpose the important set is the set of internal real fermions
$\{y,\omega\vert{\bar y},{\bar\omega}\}^{1,\cdots,6}$. We can bosonize
the fermions in this set by defining
\beq
{\rm e}^{iX_i}= {1\over\sqrt2}(y_i+i\omega_i)~~~
{\rm e}^{i{\bar X}_i}= {1\over\sqrt2}({\bar y}_i+i{\bar\omega}_i)
\label{bozonization}
\eeq
We recall that the vacuum of the sectors $b_i$ is made of 12 periodic complex
fermions, $f$, each producing a doubly degenerate vacua
${\vert +\rangle},{\vert -\rangle}$ ,
annihilated by the zero modes $f_0$ and
${{f_0}^*}$ and with fermion numbers  $F(f)=0,-1$, respectively.
The total number of states in each of these sector is therefore
$$2^{12}=\sum_{n=0}^{12}\left(\matrix{12 \cr n\cr}\right).$$
After applying the GSO projections the degeneracy at the level of the 
extended NAHE model distributes as follows:

\begin{eqnarray}
~~~~~~~~~\left[~~y,\omega\vert{\bar y}, {\bar\omega}~~\right]_{b_j}
& \left( \psi^\mu\ ,  \chi_j\right) ~~~~
  \left[~~ {\bar\psi}^{1,\cdots,5}~~\right]  ~~~~~~~~~~~~~
\left({\bar\eta}_j\right)~~\nonumber\\
\left[\left(\matrix{4\cr
                                    0\cr}\right)+
\left(\matrix{4\cr
                                    2\cr}\right)+
\left(\matrix{4\cr
                                    4\cr}\right)\right]
& \left\{ 
\left(\matrix{2\cr
                                    0\cr}\right)\right.
 {\left[ \left(\matrix{5\cr
                                    0\cr}\right)+
\left(\matrix{5\cr
                                    2\cr}\right)+   
\left(\matrix{5\cr
                                    4\cr}\right)\right]
\left(\matrix{1\cr
                                    0\cr}\right)}\nonumber\\
&+~\left(\matrix{2\cr
                                   2\cr}\right)
{\left[\left(\matrix{5\cr
                                    1\cr}\right)+
\left(\matrix{5\cr
                                    3\cr}\right)+
\left(\matrix{5\cr
                                    5\cr}\right)\right]\left.
\left(\matrix{1\cr
                                    1\cr}\right)\right\}}
\label{spinor}
\end{eqnarray}
where
$4=\{y^3y^4,y^5y^6,{\bar y}^3{\bar y}^4,
{\bar y}^5{\bar y}^6\}$, $2=\{\psi^\mu,\chi^{12}\}$,
$5=\{{\bar\psi}^{1,\cdots,5}\}$ and $1=\{{\bar\eta}^1\}$.
The combinatorial factor counts the number of $\vert{-}\rangle$ in
a given state. The two terms in the curly brackets correspond to the two
components of a Weyl spinor.  The $10+1$ in the $27$ of $E_6$ are
obtained from the sector $b_j+\xi_1$.
The states which count the multiplicities of $E_6$ are the internal
fermionic states $\{y^{3,\cdots,6}\vert{\bar y}^{3,\cdots,6}\}$.
A similar result is
obtained for the sectors $b_2$ and $b_3$ with $\{y^{1,2},\omega^{5,6}
\vert{\bar y}^{1,2},{\bar\omega}^{5,6}\}$
and $\{\omega^{1,\cdots,4}\vert{\bar\omega}^{1,\cdots,4}\}$
respectively, which suggests that
these twelve states correspond to a six dimensional
compactified orbifold with Euler characteristic equal to 48.

The construction of the free fermionic models beyond the NAHE--set
entails the construction of additional boundary condition basis 
vectors and the associated one--loop GSO phases. Their
function is to reduce the number of generations and 
at the same time break the four dimensional gauge group.
In terms of the former the reduction is primarily by the
action on the set of internal world--sheet fermions
$\{y,\omega|{\bar y}, {\bar\omega}\}$.
As elaborated in the next section this 
set corresponds to the internal compactified manifold
and the action of the additional boundary condition basis
vectors on this set also breaks the gauge symmetries 
from the internal lattice enhancement. The later is 
obtained by the action on the gauge degrees of freedom
which correspond to the world--sheet fermions
$\{{\bar\psi}^{1,\cdots,5},{\bar\eta}^{1,\cdots,3},{\bar\phi}^{1,\cdots,8}\}$.
In the bosonic formulation this would correspond to
Wilson--line breaking of the gauge symmetries, hence for the purpose
of the reduction of the number of generations we
can focus on the assignment to the internal world--sheet fermions
$\{y,\omega|{\bar y},{\bar\omega}\}$.

We can therefore examine basis vectors that do not break the
gauge symmetries further, {\it i.e.} basis vectors of the 
form $b_j$, with 
$$\{\psi^\mu_{1,2}\chi_{j,j+1},(y,\omega|{\bar y},
{\bar\omega}),{\bar\psi}^{1,\cdots,5},{\bar\eta}_j\}=1$$
for some selection of $(y,\omega|{\bar y}, {\bar\omega})=1$
assignments such that the additional vectors
$b_j$ produce massless $SO(10)$ spinorials. We will refer
to such vectors as spinorial vectors. The additional
basis vectors $b_j$ can then produce chiral, or non--chiral, spectrum. 
The condition that the spectrum from a given such sector $b_j$
be chiral is that there exist another spinorial
vector, $b_i$, in the additive
group $\Xi$, such that the overlap between the
periodic fermions of the internal set $(y,\omega|{\bar y},
{\bar\omega})$ is empty, {\it i.e.}
\begin{equation}
\{b_j(y,\omega|{\bar y},{\bar\omega})\}\cap
\{b_i(y,\omega|{\bar y},{\bar\omega})\}\equiv\emptyset~.
\label{chiralitycondition}
\end{equation}
If there exists such a vector $b_i$ in the additive group then it
will induce a GSO projection that will select
the chiral states from the sector $b_j$. Interchangeably,
if such a vector does not exist, the states from the sector
$b_j$ will be non--chiral, ${\it i.e.}$ there will be an equal
number of $16$ and $\overline{16}$ or $27$ and $\overline{27}$.
For example, we note that for the NAHE--set basis vectors
the condition (\ref{chiralitycondition}) is satisfied.
Below I discuss the orbifold
correspondence of this condition.
The reduction to three generations in a specific model
is illustrated in table \ref{m278}.

\begin{eqnarray}
 &\begin{tabular}{c|c|ccc|c|ccc|c}
 ~ & $\psi^\mu$ & $\chi^{12}$ & $\chi^{34}$ & $\chi^{56}$ &
        $\bar{\psi}^{1,...,5} $ &
        $\bar{\eta}^1 $&
        $\bar{\eta}^2 $&
        $\bar{\eta}^3 $&
        $\bar{\phi}^{1,...,8} $\\
\hline
\hline
  ${\alpha}$  &  0 & 0&0&0 & 1~1~1~0~0 & 0 & 0 & 0 & 1~1~1~1~0~0~0~0 \\
  ${\beta}$   &  0 & 0&0&0 & 1~1~1~0~0 & 0 & 0 & 0 & 1~1~1~1~0~0~0~0 \\
  ${\gamma}$  &  0 & 0&0&0 &
		${1\over2}$~${1\over2}$~${1\over2}$~${1\over2}$~${1\over2}$
	      & ${1\over2}$ & ${1\over2}$ & ${1\over2}$ &
                ${1\over2}$~0~1~1~${1\over2}$~${1\over2}$~${1\over2}$~0 \\
\end{tabular}
   \nonumber\\
   ~  &  ~ \nonumber\\
   ~  &  ~ \nonumber\\
     &\begin{tabular}{c|c|c|c}
 ~&   $y^3{y}^6$
      $y^4{\bar y}^4$
      $y^5{\bar y}^5$
      ${\bar y}^3{\bar y}^6$
  &   $y^1{\omega}^5$
      $y^2{\bar y}^2$
      $\omega^6{\bar\omega}^6$
      ${\bar y}^1{\bar\omega}^5$
  &   $\omega^2{\omega}^4$
      $\omega^1{\bar\omega}^1$
      $\omega^3{\bar\omega}^3$
      ${\bar\omega}^2{\bar\omega}^4$ \\
\hline
\hline
$\alpha$ & 1 ~~~ 0 ~~~ 0 ~~~ 0  & 0 ~~~ 0 ~~~ 1 ~~~ 1  & 0 ~~~ 0 ~~~ 1 ~~~ 1
\\
$\beta$  & 0 ~~~ 0 ~~~ 1 ~~~ 1  & 1 ~~~ 0 ~~~ 0 ~~~ 0  & 0 ~~~ 1 ~~~ 0 ~~~ 1
\\
$\gamma$ & 0 ~~~ 1 ~~~ 0 ~~~ 1  & 0 ~~~ 1 ~~~ 0 ~~~ 1  & 1 ~~~ 0 ~~~ 0 ~~~ 0
\\
\end{tabular}
\label{m278}
\end{eqnarray}
In the realistic free fermionic models the vector $X$
is replaced by the vector $2\gamma$ in which $\{{\bar\psi}^{1,\cdots,5},
{\bar\eta}^1,{\bar\eta}^2,{\bar\eta}^3,{\bar\phi}^{1,\cdots,4}\}$
are periodic. This reflects the fact that these models
have (2,0) rather than (2,2) world-sheet supersymmetry.
At the level of the NAHE set we have 48 generations.
One half of the generations is projected because of the vector $2\gamma$.
Each of the three vectors in table \ref{m278}
acts nontrivially on the degenerate
vacuum of the fermionic states
$\{y,\omega\vert{\bar y},{\bar\omega}\}$ that are periodic in the
sectors $b_1$, $b_2$ and $b_3$ and reduces the combinatorial
factor of Eq. (\ref{spinor}) by a half.
Thus, we obtain one generation from each sector $b_1$, $b_2$ and $b_3$.

The geometrical interpretation of the basis vectors
beyond the NAHE set is facilitated by taking combinations of the
basis vectors in \ref{m278}, which entails choosing another set
to generate the same vacuum. The combinations
$\alpha+\beta$, $\alpha+\gamma$, $\alpha+\beta+\gamma$ produce
the following boundary conditions under the set of internal
real fermions

\begin{eqnarray}
     &\begin{tabular}{c|c|c|c}
 ~&   $y^3{y}^6$
      $y^4{\bar y}^4$
      $y^5{\bar y}^5$
      ${\bar y}^3{\bar y}^6$
  &   $y^1{\omega}^5$
      $y^2{\bar y}^2$
      $\omega^6{\bar\omega}^6$
      ${\bar y}^1{\bar\omega}^5$
  &   $\omega^2{\omega}^4$
      $\omega^1{\bar\omega}^1$
      $\omega^3{\bar\omega}^3$
      ${\bar\omega}^2{\bar\omega}^4$ \\
\hline
\hline
$\alpha+\beta$ 
& 1 ~~~ 0 ~~~ 1 ~~~ 1  & 1 ~~~ 0 ~~~ 1 ~~~ 1  & 0 ~~~ 1 ~~~ 1 ~~~ 0
\\
$\beta+\gamma$
& 0 ~~~ 1 ~~~ 1 ~~~ 0  & 1 ~~~ 1 ~~~ 0 ~~~ 1  & 1 ~~~ 1 ~~~ 0 ~~~ 1
\\
$\alpha+\beta+\gamma$
& 1 ~~~ 1 ~~~ 1 ~~~ 0  & 1 ~~~ 1 ~~~ 1 ~~~ 0  & 1 ~~~ 1 ~~~ 1 ~~~ 0
\\
\\
\end{tabular}
\label{m2782}
\end{eqnarray}

It is noted that the two combinations $\alpha+\beta$ and $\beta+\gamma$
are fully symmetric between the left and right movers, whereas the
third, $\alpha+\beta+\gamma$, is fully asymmetric. From eq.
(\ref{bozonization}) we note that the action of the first two combinations
on the compactified bosonic coordinates translates therefore to symmetric 
shifts. Thus, we see that reduction of the number of generations
is obtained by further action of fully symmetric shifts.

Due to the presence of the third combination the situation, however, 
is more complicated. The third combination in \ref{m2782} is fully 
asymmetric between the left and right movers and therefore
does not have an obvious geometrical interpretation.
Three generations are obtained in the free fermionic models
by the inclusion of the asymmetric shift. This observation
has profound implications on the type of geometries that
are related to the realistic string vacua, as well as on the
issue of moduli stabilization.

\section{Bosonic classification}

It is instructive to classify all possible quotients of the $Z_2\times Z_2$
orbifold by additional symmetric shifts of order two on the three complex
tori.  Starting with three complex tori parameterized by three
complex coordinates, the torus identification is given by (\ref{t2cube}). 
The symmetric shift actions are 
$$z_i=z_i+{1\over 2}~~~{\rm and}~~~z_i=z_i+{\tau\over2}$$
and a given action may act on any number of the three tori.
The additional shifts may have the following actions:
\begin{eqnarray}
    {\rm freely ~acting}  & ~~~~~\longrightarrow \oplus (h_{11}=h_{21}=0) \\
{\rm chiral ~preserving}  & \longrightarrow \oplus (h_{11}=h_{21}) \\
{\rm non ~freely ~acting} & \longrightarrow \oplus (h_{11}\ne h_{21})
\end{eqnarray}

In the first case one of the tori is always shifted and hence there
are no fixed points and the action is free. In the second case
we have tori above fixed points and all the other geometrical
identifications preserve the fixed tori. Since the contribution
of $T_2$ gives $\oplus h_{11}=\oplus h_{21}=1$
we have that this case preserves the chirality. In the third case
we have a situation that for a fixed torus we impose the identification
$z_k \leftrightarrow -z_k$. In this case the torus above the fixed
point degenerates to $P_1$, for which $\oplus h_{11}=1$,
$\oplus h_{21}=0$
and therefore this case adds to the net chirality. 
In ref. \cite{ron} we have classified all
the possible shifts on the three complex tori.
The outcome of this classification is that quotients
of the original $Z_2\times Z_2$ orbifold solely by symmetric
shifts on the three internal $T_2$ tori do not produce
a manifold with cohomology that corresponds to three generation.
The analysis indicates that three generations are not possible for
$Z_2\times Z_2$ orbifolds of three complex tori, with purely symmetric shifts.

Thus, the reduction to three generations seems to necessitate
an operation, which is asymmetric between the left-- or the right--movers. 
One possibility is the asymmetric orbifold that operates in the
case of the realistic free fermionic models. Another option
may be to utilize the Wilson line breaking of the
four dimensional gauge group \cite{nilles}.
In the first case the incorporation of an asymmetric shift
in the reduction to three generations, has profound implications for the
issues of moduli stabilization and vacuum selection. The reason
being that it can only be implemented at enhanced symmetry
points in the moduli space. In this context we envision again that the
self--dual point under T--duality plays a special role. In the 
context of nonperturbative dualities the dilaton and
moduli are interchanged, with potentially important
implications for the problem of dilaton stabilization.

To summarize this section, the argument here is that T--duality
is the pivotal property of string theory in trying to understand
the vacuum selection mechanism. In this context the self--dual points
may play an important role. It is then extremely intriguing that
precisely in the vicinity of the self--dual point there exist 
a class of models that capture the two main characteristics
of the Standard Model. The existence of three generations
together with their $SO(10)$ embedding.

\section{Fermionic classification}

The classification of the $Z_2\times Z_2$ orbifold can proceed in
a systematic fashion by utilizing the free fermionic formalism
\cite{fff,kounnas}.
The partition function of a generic $N=1$ $(2,2)$ supersymmetric 
heterotic string vacuum at a generic point in the moduli space
is schematically given by 
\beq
Z={1\over{\tau_2}}{1\over{|\eta|^4}}
{{\rm e}^{i\pi\phi_L}\over{\eta^{10}{\bar\eta^{22}}}}
Z^F_L\Gamma_{6,6} Z_R^G \Gamma_{0,13},
\label{neqoneparti}
\eeq
where $\phi_L$ is a phase that depends on the assignment of 
boundary conditions for the world--sheet fermions,
$Z_L^F$ is the term accounting for the left--moving fermionic
superpartners; $Z_R^G$ are the corresponding degrees of freedom
on the right--moving side that reflect the $(2,2)$ world--sheet
supersymmetric structure; $\Gamma_{0,13}$ account for the remaining
gauge degrees of freedom on the non--supersymmetric side;
and finally $\Gamma_{6,6}$ accounts for the left--right
symmetric degrees of freedom that correspond to the
six dimensional compactified manifolds. In a bosonic
formalism this segment of the partition function incorporates
the dependence on the moduli, and typically factorizes into
product of three $T_2$ tori, or six $S_1$ circles, {\it i.e.}
\beq
\Gamma_{6,6}~\rightarrow~\Gamma_{2,2}^3~\rightarrow~\Gamma_{1,1}^6~.
\label{g66g22g11}
\eeq
Now, the point is that, as we have discussed in section \ref{post},
the point in the moduli space at which the internal dimensions
can be represented as free fermions propagating on the string
world--sheet, corresponds to fixing the moduli that appear in
$\Gamma_{6,6}$ at some specific value. As discussed in section
\ref{post}, the free fermionic point of a six dimensional
toroidal lattice corresponds to the point of maximally enhanced
symmetry point, or to a self--dual point up to  to a
$SL(D,Z)$ and $\Theta(Z)$ transformations \cite{grv,tduality}.
That is we have the result that fixing the moduli of $\Gamma_{6,6}$
at this ``self--dual point'' reproduces the partition function
at the free fermionic point, {\it i.e.}
\beq
\Gamma_{6,6}(``{\rm self-dual~point}'')~=~
\Gamma_{6,6}({\rm free~fermionic~point})~.
\label{gsdgffp}
\eeq
The crucial point is that the chiral spectrum of the $N=1$
string vacuum that arise from the twisted sector is independent
of the moduli, and the untwisted sector of the $Z_2\times Z_2$ orbifold
always contributes $(h_{11},h_{21})=(3,3)$. This fact allows us
to use the free fermionic tools to completely classify the class
of $Z_2\times Z_2$ orbifold compactifications by their chiral content.
We can then reincorporate the moduli dependence through the relation
(\ref{gsdgffp}), to completely classify the $Z_2\times Z_2$ orbifolds
at generic points in the moduli space. While the partition function
in eq. (\ref{neqoneparti}) alluded to $(2,2)$ world--sheet supersymmetry,
in fact the class of models that can be classified is more general,
and the right--moving world--sheet supersymmetry may be broken
by GSO phases.

The fermionic methods entail choosing a set of boundary condition
basis vectors and one--loop GSO projection coefficients.
In the free fermionic formulation the 4-dimensional heterotic string,
in the light-cone gauge, is described
by $20$ left--moving  and $44$ right--moving real fermions.
A large number of models can be constructed by choosing
different phases picked up by   fermions ($f_A, A=1,\dots,44$)
when transported along
the torus non-contractible loops.
Each model corresponds to a particular choice of fermion phases
consistent with modular invariance
that can be generated by a set of  basis vectors $v_i,i=1,\dots,n$
$$v_i=\left\{\alpha_i(f_1),\alpha_i(f_{2}),\alpha_i(f_{3}))\dots\right\}$$
describing the transformation  properties of each fermion
\begin{equation}
f_A\to -e^{i\pi\alpha_i(f_A)}\ f_A, \ , A=1,\dots,44
\end{equation}
The basis vectors span a space $\Xi$ which consists of $2^N$
sectors that give rise to the string spectrum. Each sector is given by
\begin{equation}
\xi = \sum N_i v_i,\ \  N_i =0,1
\end{equation}
The spectrum is truncated by a generalized GSO projection whose
action on a string state  $|S>$ is
\begin{equation}\label{eq:gso}
e^{i\pi v_i\cdot F_S} |S> = \delta_{S}\ \cc{S}{v_i} |S>,
\end{equation}
where $F_S$ is the fermion number operator and $\delta_{S}=\pm1$
is the spacetime spin statistics index.
Different sets of projection coefficients $\cc{S}{v_i}=\pm1$
consistent with modular invariance give
rise to different models. Summarizing: a model can be defined
uniquely by a set of basis vectors $v_i,i=1,\dots,n$
and a set of $2^{N(N-1)/2}$ independent projections coefficients
$\cc{v_i}{v_j}, i>j$.

\subsection{General setup}
The free fermions in the light-cone gauge in the usual notation are:
$\psi^\mu, \chi^i,y^i, \omega^i, i=1,\dots,6$ (left-movers) and
$\bar{y}^i,\bar{\omega}^i, i=1,\dots,6$,
$\psi^A, A=1,\dots,5$, $\bar{\eta}^B, B=1,2,3$,
$\bar{\phi}^\alpha, \alpha=1,\ldots,8$ (right-movers).
The class of models we investigate, is generated by a set
of 12 basis vectors
$
B=\{v_1,v_2,\dots,v_{12}\},~
$
where
\begin{eqnarray}
v_1=1&=&\{\psi^\mu,\
\chi^{1,\dots,6},y^{1,\dots,6},\omega^{1,\dots,6}|\bar{y}^{1,\dots,6},
\bar{\omega}^{1,\dots,6},\bar{\eta}^{1,2,3},
\bar{\psi}^{1,\dots,5},\bar{\phi}^{1,\dots,8}\},\nn\\
v_2=S&=&\{\psi^\mu,\chi^{1,\dots,6}\},\nn\\
v_{2+i}=e_i&=&\{y^{i},\omega^{i}|\bar{y}^i,\bar{\omega}^i\}, \ i=1,\dots,6,\nn\\
v_{9}=b_1&=&\{\chi^{34},\chi^{56},y^{34},y^{56}|\bar{y}^{34},
\bar{y}^{56},\bar{\eta}^1,\bar{\psi}^{1,\dots,5}\},\label{basis}\\
v_{10}=b_2&=&\{\chi^{12},\chi^{56},y^{12},y^{56}|\bar{y}^{12},
\bar{y}^{56},\bar{\eta}^2,\bar{\psi}^{1,\dots,5}\},\nn\\
v_{11}=z_1&=&\{\bar{\phi}^{1,\dots,4}\},\nn\\
v_{12}=z_2&=&\{\bar{\phi}^{5,\dots,8}\}.\nn
\end{eqnarray}
The vectors $1,S$ generate an
$N=4$ supersymmetric model. The vectors $e_i,i=1,\dots,6$ give rise
to all possible symmetric shifts of internal fermions
($y^i,\omega^i,\bar{y}^i,\bar{\omega}^i$) while $b_1$ and $b_2$
represent  the $Z_2\times Z_2$ orbifold twists. The remaining fermions
not affected by the action
of the previous vectors are $\bar{\phi}^i,i=1,\dots,8$ which
normally give rise to the hidden sector gauge group.
The vectors $z_1,z_2$ divide these eight fermions into two sets of
four which in the $Z_2\times{Z_2}$ case is the maximum consistent
partition\cite{fff}.
This is the most general basis, with symmetric shifts for the internal
fermions, that is compatible with a
Kac--Moody level one $SO(10)$ embedding.
Without loss of generality we can set the associated projection coefficients
\begin{eqnarray}
\cc{1}{1}=\cc{1}{S}=\cc{S}{S}=\cc{S}{e_i}=\cc{S}{b_A}=-\cc{b_2}{S}=\cc{S}{z_n}=-1,
\label{asum}\end{eqnarray}
leaving  55 independent coefficients
\begin{eqnarray}
&&\cc{e_i}{e_j}, i\ge j, \ \ \cc{b_1}{b_2}, \ \ \cc{z_1}{z_2},\nn\\
&&\cc{e_i}{z_n}, \cc{e_i}{b_A},\cc{b_A}{z_n},
\ \ i,j=1,\dots6\,\ ,\  A,B,m,n=1,2\nn.
\end{eqnarray}
The remaining  projection coefficients are determined by modular
invariance \cite{fff}. Each of the linearly independent coefficients
can take two discrete values $\pm1$ and thus a simple counting
gives  $2^{55}$ (that is approximately $10^{16.6}$) distinct models
in the class under consideration.

\subsection{The analysis}
In a generic model described above the gauge group has the form
$$
SO(10)\times{U(1)}^3\times{SO(8)}^2
$$
Depending on the  choices of the projection coefficients,
extra gauge bosons arise from
$
x=1+S+\sum_{i=1}^{6}e_i+z_1+z_2=\{{\bar{\eta}^{123},\bar{\psi}^{12345}}\}
$
resulting in the enhancement $SO(10)\times{U(1)}\to E_6$.
Additional gauge bosons can
arise  from the sectors
$z_1,z_2$ and $z_1+z_2$ and  enhance
${SO(8)}^2\to SO(16)$ or ${SO(8)}^2\to E_8$.
For particular choices of the projection coefficients
other gauge groups can be obtained \cite{prep}.

The untwisted sector matter is common to all models
(putting aside gauge group enhancements)
and consists of six vectors of $SO(10)$ and
12 non-Abelian gauge group singlets.
Chiral twisted matter arises from the following 48 sectors
(16 per orbifold plane)
\begin{eqnarray}
B_{\ell_3^1\ell_4^1\ell_5^1\ell_6^1}^1&=&S+b_1+\ell_3^1 e_3+\ell_4^1 e_4 +
\ell_5^1 e_5 + \ell_6^1 e_6 \nn\\
B_{\ell_1^2\ell_2^2\ell_5^2\ell_6^2}^2&=&S+b_2+\ell_1^2 e_1+\ell_2^2 e_2 +
\ell_5^2 e_5 + \ell_6^2 e_6 \label{ss}\\
B_{\ell_1^3\ell_2^3\ell_3^3\ell_4^3}^3&=&
S+b_3+ \ell_1^3 e_1+\ell_2^3 e_2 +\ell_3^3 e_3+ \ell_4^3 e_4\nn
\end{eqnarray}
where $\ell_i^j=0,1$ and $b_3=1+S+b_1+b_2+\sum_{i=1}^6 e_i+\sum_{n=1}^2 z_n$.
These states are  spinorials of $SO(10)$ and one can obtain at maximum one
spinorial ($\bf 16$ or
$\bf {\overline{{16}}}$) per sector and thus totally 48 spinorials.
Extra non chiral matter i.e. vectors of $SO(10)$ as well as
singlets arise from the sectors
$S+b_i+b_j+e_m+e_n$.

In our formulation we have separated the spinorials, that is we
have separated the 48 fixed points
of the $Z_2\times{Z_2}$ orbifold. This separation allows us to
examine the GSO action,
depending on the projection coefficients, on each spinorial separately.
The choice of these coefficients
determines which  spinorials are  projected out, as well as
the chirality of the surviving states.

One of the  advantages of our formulation is that it allows to extract generic
formulas regarding the
number and the chirality of each spinorial.
This is important because it allows an algebraic
treatment of the entire class of models without
deriving each model explicitly.
The number of surviving spinorials per sector (\ref{ss}) is given by
\begin{eqnarray}
P_{\ell_3^1\ell_4^1\ell_5^1\ell_6^1}^{(1)}&=&
\frac{1}{16}\,\prod_{i=1,2}\left(1-\cc{e_i}
{B_{\ell_3^1\ell_4^1\ell_5^1\ell_6^1}^{(1)}}\right)\,
\prod_{m=1,2}\left(1-\cc{z_i}
{B_{\ell_3^1\ell_4^1\ell_5^1\ell_6^1}^{(1)}}\right)\,\label{pa}\\
P_{\ell_1^2\ell_2^2\ell_5^2\ell_6^2}^{(2)}&=&
\frac{1}{16}\,\prod_{i=3,4}\left(1-\cc{e_i}
{B_{\ell_1^2\ell_2^2\ell_5^2\ell_6^2}^{(2)}}\right)\,
\prod_{m=1,2}\left(1-\cc{z_m}
{B_{\ell_1^2\ell_2^2\ell_5^2\ell_6^2}^{(2)}}\right)\,\label{pb}
\\
P_{\ell_1^3\ell_2^3\ell_3^3\ell_4^3}^{(3)}&=&
\frac{1}{16}\,\prod_{i=5,6}\left(1-\cc{e_i}
{B_{\ell_1^3\ell_2^3\ell_3^3\ell_4^3}^{(3)}}\right)\,
\prod_{m=1,2}\,\left(1-\cc{z_m}
{B_{\ell_1^3\ell_2^3\ell_3^3\ell_4^3}^{(3)}}\right)\,\label{pc}
\end{eqnarray}
and thus the total number of spinorial per model is the sum of the above.
The chirality of the surviving spinorials is given by
\begin{equation}
X_{\ell_3^1\ell_4^1\ell_5^1\ell_6^1}^{(1)}=
\cc{b_2+(1-\ell_5^1) e_5+(1-\ell_6^1) e_6}
{B_{\ell_3^1\ell_4^1\ell_5^1\ell_6^1}^{(1)}}\label{ca}
\end{equation}
\begin{equation}
X_{\ell_1^2\ell_2^2\ell_5^2\ell_6^2}^{(2)}=
\cc{b_1+(1-\ell_5^2) e_5+(1-\ell_6^2) e_6}
{B_{\ell_1^2\ell_2^2\ell_5^2\ell_6^2}^{(2)}}\label{cb}
\end{equation}
\begin{equation}
X_{\ell_1^3\ell_2^3\ell_3^3\ell_4^3}^{(3)}=
\cc{b_1+(1-\ell_3^3) e_3+(1-\ell_4^3) e_4}
{B_{\ell_1^3\ell_2^3\ell_3^3\ell_4^3}^{(3)}}\label{cc}
\end{equation}
The net number of families is then
\begin{equation}
N=-\sum_{I=1}^3\sum_{p,q,r,s=0}^1X^{(I)}_{pqrs} P_{pqrs}^{(I)}
\end{equation}
Similar formulas can be easily derived for the number of 
vectorials and the number of singlets and can be extended
to the $U(1)$ charges.

Formulas (\ref{pa})-(\ref{pc}) allow us to identify the mechanism of
spinorial reduction, or in other words
the fixed point reduction, in the fermionic language.
For a particular sector ($B_{pqrs}$) of the orbifold plane $i$
there exist two shift vectors ($e_{2i-1}, e_{2i}$)
and the two zeta vectors ($z_1,z_2$) that have no common
elements with $B_{pqrs}$. Setting the relative projection
coefficients (\ref{pc})
to $-1$, each of the above four vectors acts as a projector
that cuts the number of fixed points in the
associated sector by a factor of two. Since four such projectors
are available for each sector the number of fixed points
can be reduced from 16 to one per plane.

The classification in the fermionic formulation therefore reduces
to scanning the range of choices for the GSO projection coefficients
and determining the net chirality for each choice. A priori, the basis
given by eq. (\ref{basis}) can produce spinorial representations
from each one of the three twisted planes. We dub such vacua as
$S^3$ models. In ref. \cite{fknr} we classified only this class of models,
which entails imposing further restriction on the one--loop GSO projection
coefficients \cite{fknr}. Other possibilities include the $S^2V$, $SV^2$ and
$V^3$ models. In the first of those two of the twisted sectors
produce spinorial representations, whereas the third produces
vectorials, and in an apparent notation for the two other cases.
A priori, we can envision producing these additional classes by
modifying the basis vectors in eq. (\ref{basis}) \cite{nooij}. However,
it turns out that the same space of models can be scanned by working with
the original basis (\ref{basis}), and modifying the GSO projection
coefficients \cite{prep}. Below I summarize the results of the
classification that was done for the class of $S^3$ models
in ref. \cite{fknr}.

\section{Results}

The main results of the classification are as follows \cite{fknr}.

\begin{itemize}
\item There exist a class of three generation models. In this
class of models the internal $\Gamma_{6,6}$ lattice is factorized
to a product of six circles, {\it i.e.}
$$\Gamma_{6,6}~\rightarrow~\Gamma_{1,1}^6.$$
In this class of models the $SO(10)$ symmetry cannot be broken
perturbatively by using Wilson lines. The reason being that the
Wilson lines breaking also projects out sub--components of the
spinorial 16 of $SO(10)$ and the resulting spectrum does not 
contain the full Standard Model matter content. Nevertheless,
one cannot exclude the possibility that some, yet unknown, nonperturbative
mechanism will allow for $SO(10)$ breaking, while keeping the full
Standard Model matter spectrum. Clearly, however, this class
of models is not amenable to perturbative analysis, and at
present is not phenomenologically viable. 

\item There does not exist in the space of vacua scanned by this
classification a three generation model in which the complex structure
is preserved, {\it i.e.} in which $\Gamma_{6,6}~\rightarrow~\Gamma_{2,2}^3.$
This result seems to indicate that there does not exist a $Z_2\times Z_2$
Calabi--Yau manifold whose cohomology correspond to a net number
of three generations, and that the perturbative three generation
$Z_2\times Z_2$ orbifolds necessarily employs an asymmetric
shift to achieve the reduction to three generations. 

\item There exist a class of models that admits an $N=4$
interpretation. This class of vacua is obtained with
the additional restricted GSO phases
\begin{equation}
\cc{b_i}{z_m}=\cc{b_i}{e_i}=+1
\end{equation}
In this class of models all the information on the chiral content
of the models is already contained in the toroidal lattice of
the ascendant $N=4$ theory. This is similar to the case of the
$Z_2\times Z_2$ orbifold of an $SO(12)$ lattice versus the 
$Z_2\times Z_2$ orbifold of an $SO(4)^3$ lattice. As discussed in
section \ref{z2z2orbifold} the first case produces 24 generations
whereas the second produces 48 generations. Thus, the chiral content
of the models is already predetermined by the $N=4$ lattice.
In this class of vacua the one--loop GSO projection coefficients
that appear in the $N=1$ partition function and determine the
number of generations, admit the interpretation
of corresponding to the fixed VEVs of the $N=4$ background fields,
whose dynamical components are projected out by the $Z_2\times Z_2$
orbifold projections.  
\item The need to use an asymmetric projection in the reduction
to three generations implies that the moduli of the internal
dimensions are fixed in the vicinity of the self--dual point.
The reason being that the asymmetric shift can only be employed
at the enhanced symmetry point, and its application projects out the moduli
fields. Hence, the utilization of the asymmetric shift implies that
the untwisted geometrical moduli are fixed in the vicinity of
the self--dual point \footnote{{\it i.e.} there may exist a factor of
$\sqrt2$ due to the mismatch between the free fermionic point
and the self--dual point. But, clearly, the moduli are fixed
at a scale which is of the order of the self--dual point.}
\end{itemize}
It ought to be remarked that the necessity to use an asymmetric
shift in the reduction to three generations in $Z_2\times Z_2$
orbifold is under dispute \cite{nilles}. In ref. \cite{nilles}
the authors utilize an alternative method of reducing the 
number of families by the Wilson line breaking of the
hidden gauge degrees of freedom and find three generation
models that do preserve the complex structure.
While the necessity to include an asymmetric shift is valid
for the class of models scanned in ref. \cite{fknr,nooij,prep},
pending an understanding of the overlap of the two methodologies
employed in ref. \cite{fknr} and \cite{nilles}, we may conclude
that, while the asymmetric shift may not be necessary to go
down to three generations, it is certainly sufficient.
As already implied above, the utilization of an asymmetric
shift has important consequences that we further expand upon
below. 

\section{Implications from S--duality?}

The utilization of the asymmetric shift in the three generation free fermionic
models implies that all the geometrical untwisted moduli are fixed in
these models. The reason being that that the asymmetric shift can only
operate at the enhanced symmetry points in the moduli space, and
that its action projects out the untwisted fields that correspond
to the geometrical moduli. Hence, the asymmetric shift acts
as a moduli fixing mechanism in the perturbative string models.
It is intriguing, but perhaps not surprising, that this mechanism
operates in three generation models, in which we anticipate that the
number of moduli is reduced. It should be remarked, however, that
there may still exist unfixed moduli in the models. These may come
from twisted moduli, that might be interchanged with the untwisted
moduli, and render the analysis more cumbersome. Additionally, of course,
the dilaton VEV remains unfixed in these perturbative models.

\begin{figure}[!h]
\centerline{\epsfxsize 3.0 truein \epsfbox {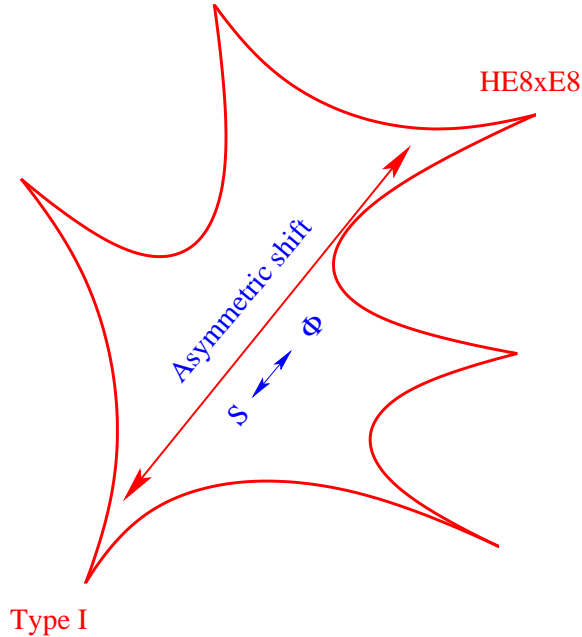}}
\caption{Implications from S--duality?}  
\label{sduality}
\end{figure}

In this respect it is intriguing to consider what the incorporation of the
asymmetric shift in the framework of M--theory dualities may imply.
Under the strong--weak duality exchange between the heterotic--string 
and type I string, the dilaton is interchanged with a moduli. Hence,
the possible implication of the asymmetric shift is that in the dual
type I picture the dilaton VEV is fixed in the vicinity of the
self--dual point. Naturally, this is an intriguing possibility, and
hopefully it can be substantiated. 

\section{Back to T--duality}

As discussed in section \ref{post} a key property of string theory
in compact space is the symmetry under exchange of a radius with
its inverse, which is accompanied with exchange of momenta and
winding modes. Reverting back to the wave--function of a
free point particle in one dimension,
$$\Psi~\sim~{\rm e}^{iP\cdot X}~,$$
we note that it is invariant under the exchange
$$P~\leftrightarrow~X.$$ However, in ordinary 
Kaluza--Klein compactification on a circle, we note that
the invariance is lost due to the quantization of the momenta
modes, $$P={m\over R}.$$
In the case of the string, as discussed in section \ref{post},
we have that, $$P~\sim~{m\over R}~+~ n R.$$
If we think of the momenta and winding modes as
the phase--space of the compact space, we have that the
introduction of string winding modes restores this invariance.
We can thus turn the table around and hypothesize that the key
physical property that string theory enables is the restoration
of the phase--space duality. Thus, the key physical property
that should underly the formalism is the requirement of manifest
phase--space duality. Furthermore, the phenomenological success
of the free fermionic models, and their association with the
self--dual points under T--duality, points to the possible
association of the vacuum state of the theory with the self--dual
state under phase--space duality. This view suggests a new starting
point for the formulation of string theory and quantum gravity,
and a constructive way to determine its vacuum. In the
following I will describe the preliminary steps in such a program.

\section{Phase--space duality}

Duality and self--duality play a key role in the recent
formulation of quantum mechanics from an equivalence
postulate \cite{fm}.
The duality in this context is phase--space duality,
which is manifested due to the involutive nature of
the Legendre transformation.
Here I would like to demonstrate the association of
the self--dual states under the phase--space duality
with the states of vanishing kinetic and potential energy.
Hence, again we note the relation between the self--duality
criteria and the trivial state of the theory. 

An instructive starting point to discuss the equivalence
postulate approach to quantum mechanics is the classical
Hamilton--Jacobi formalism. The classical Hamilton equations
of motion
\beq
\dot q={\partial H\over\partial p},\qquad\dot p=-{\partial H\over\partial q},
\label{heqofm}
\eeq
are invariant under the transformations $p\rightarrow q$, $q \rightarrow -p$.
However, in general in classical mechanics this duality is lost
when a classical potential is specified. A viable question
is therefore whether one can formulate classical mechanics
with manifest phase--space duality.
It turns out that phase--space duality cannot hold for all physical classical
systems, and the break down is precisely for the states with vanishing
energy and vanishing potential. The requirement that the phase--space duality
holds for all physical system necessitates the quantum modification
of classical mechanics \cite{fm}.
In the Hamilton--Jacobi formalism of classical mechanics
the phase--space variables are related by Hamilton's generating
function $p=\partial_q{S}_0(q)$. With the new generating function
${ T}_0(p)$ defined by $q=\partial_p{ T}_0$, the two generating
functions are related  by the dual Legendre transformations \cite{fm},
$$
{ S}_0=p\partial_p{ T}_0-{ T}_0
$$
and
$$
{ T}_0=q\partial_q{ S}_0-{S}_0.
$$  
Each of the Legendre transformations are related to a second
differential equation \cite{fm} given in (\ref{2ndorderdifeq}).
Thus, we obtain the two dual pictures,

\beqn
p = {{\partial S_0}\over{\partial q}}~~~~~~~~~~~~~~~~~~~~~~~&, &~~~~~~~~~~~~~~
q={{\partial T_0}\over{\partial p}},~~~~~~~~~~~~~~~~~~~~\\
S_0= p{{\partial T_0\over\partial p}}- T_0~~~~~~~~~~~~~~~&, &~~~~~~~~~~~~~
T_0=q{{\partial S_0\over\partial q}}- S_0,\label{legendreduals}\\
\left({{\partial^2~~}\over\partial S_0^2} + U(S_0)\right)
\left({{q\sqrt p}\atop \sqrt{p}}\right)=0 &, & ~~~~~~~~~~
\left({{\partial^2~~}\over\partial T_0^2}+{\cal V}(T_0)\right)
\left({{p\sqrt q}\atop \sqrt{q}}\right)=0,~~~~~\label{2ndorderdifeq}
\eeqn
described by the two dual
sets of differential equations. 
Two points are important to note.
The first is that because the Legendre
transformation is not defined for linear functions, {\it i.e.} for
physical systems with ${ S}_0=A q+B$, it implies that the
Legendre duality fails for the free system and for the
free system with vanishing energy. Thus, we have the general condition
that $S_0$ is never a linear function of the coordinate {\it i.e.},
$$S_0~\ne~Aq~+~B~.$$
The second observation is that
there exist a set of solutions, labeled by $pq=\gamma$,
where $\gamma$ is a constant to be determined, which are simultaneous
solutions of the two sets of differential equations. These are the 
self dual states under the phase--space duality, which
are of the form \cite{fm},
$$S_0~\sim~ \ln q.$$

\section{The quantum equivalence postulate}

The Legendre phase--space
duality and its breakdown for the free system are intimately
related to the equivalence postulate, which states
that all physical systems labeled by the function
${ W}(q)=V(q)-E$, can be connected by a coordinate
transformation, $q^a\rightarrow q^b=q^b(q^a)$, defined
by $${ S}_0^b(q^b)={ S}_0^a(q^a).$$
This postulate implies that there
always exist a coordinate transformation connecting  
any state to the state ${ W}^0(q^0)=0$. Inversely, this means
that any physical state can be reached from the
state ${ W}^0(q^0)$ by a coordinate transformation.
This postulate cannot be consistent
with classical mechanics. The reason being that in Classical
Mechanics (CM) the state ${ W}^0(q^0)\equiv0$ remains a fixed
point under coordinate transformations. Thus, in CM it
is not possible to generate all states by a coordinate
transformation from the trivial state. Consistency of the
equivalence postulate implies the modification of CM,
which is analyzed by adding a still unknown function
$Q$ to the Classical Hamilton--Jacobi Equation (CHJE).
Consistency of the equivalence postulate fixes the
transformation properties for ${ W}(q)$,
$$
{ W}^v(q^v)=
 \left(\partial_{q^v}q^a\right)^2{ W}^a(q^a)+(q^a;q^v),
$$
and for $Q(q)$,
$$
 Q^v(q^v)=\left(\partial_{q^v}q^a\right)^2Q^a(q^a)-(q^a;q^v),
$$
which fixes the cocycle condition for the inhomogeneous term
$$
(q^a;q^c)=\left(\partial_{q^c}q^b\right)^2[(q^a;q^b)-(q^c;q^b)].
$$
The cocycle condition is invariant under M\"obius transformations
and fixes the functional form of the inhomogeneous term.
The cocycle condition is generalizable to higher, Euclidean or
Minkowski \cite{fm}, dimensions,
where the Jacobian of the coordinate transformation extends
to the ratio of momenta in the transformed and original systems\cite{fm}.
The identity   
$$
({\partial_q{ S}_0})^2=
\hbar^2/2\left(\{\exp(i2{ S}_0/\hbar,q)\}-\{{ S}_0,q\}\right),
$$
which embodies the equivalence postulate,
leads to the Schr\"odinger equation.
Making the identification
\beq
{ W}(q)= V(q) - E = -{\hbar^2/{4m}}\{{\rm e}^{(i2{S}_0/\hbar)},q\},
\eeq
and
$$
{Q}(q)=  {\hbar^2/{4m}}\{{ S}_0,q\},
$$
we have that   
${ S}_0$ is solution of the Quantum Stationary
Hamilton--Jacobi Equation (QSHJE),
\beq
({1/{2m}})\left({{\partial_q S}_0}\right)^2+
V(q)-E+({\hbar^2/{4m}})\{{ S}_0,q\}=0,
\label{qshje}
\eeq
where $\{,\}$ denotes the Schwarzian derivative.
{}From the identity we deduce that the trivializing
map is given by $q\rightarrow {\tilde q}=\psi^D/\psi$,
where $\psi^D$ and $\psi$ are the two linearly independent
solutions of the corresponding Schr\"odinger equation \cite{fm,floyd}.
We see that the consistency of the equivalence postulate
forces the appearance of quantum mechanics and
of $\hbar$ as a covariantizing parameter.

\section{The role of the self--dual states}
The remarkable property of the QSHJE, which distinguishes
it from the classical case, is that it admits non--trivial solution
also for the trivial state, ${ W}(q)\equiv0$.
Classical phase--space is described by the
Classical Stationary Hamilton--Jacobi Equation (CSHJE)
$$
({1/{2m}})\left({{\partial_q S}_0}\right)^2+
V(q)-E=0. 
$$
The trivial solution, with
\beq
V(q)=0~~,~~E=0,
\label{vacuumstate}
\eeq
is given by ,
$$S_0~=Aq+B,$$
{\it i.e.} precisely the solution which is not compatible with the
Legendre duality transformation, which is not defined for linear
functions. This solution is also incompatible with the equivalence
postulate. On the other hand in the case of the quantum phase--space,
which is described by the QSHJE (\ref{qshje}), the state (\ref{vacuumstate}),
admits a non--trivial solution. 
In fact the QSHJE implies that ${ S}_0=constant$ is
not an allowed solution. The fundamental characteristic
of quantum mechanics in this approach is that ${ S}_0\ne Aq+B$.
Rather, the solution for the trivial state, with $V(q)=0$ and $E=0$,
is given by
$$
{ S}_0=i\hbar/2\ln q,
$$
up to M\"obius transformations. Remarkably, this quantum
ground state solution coincides with the self--dual state
of the Legendre phase--space transformation and its dual.
Thus, we have that the quantum self--dual state plays a pivotal
role in ensuring both the consistency of the equivalence
postulate and definability of the Legendre phase--space  
duality for all physical states. The association of the
self--dual state and the physical state with $V(q)=0$ and
$E=0$ suggests a criteria by which the vacuum state of a given
physical system could be identified. Namely, if one can
identify correctly the complete phase--space and its duality
structure, the vacuum state will then be identified with the
self--dual states. Note also that the fact that the quantum potential
is never vanishing, implies that even the trivial quantum state
has a non--vanishing quantum potential. 

\section{Conclusions}

String theory is in a precarious state of affairs. On the one hand the theory
clearly exhibits great promise in providing a consistent framework for
quantum gravity. On the other the need to embed the theory in 
higher dimensions is troubling. Furthermore, the apparent existence
of a multitude of vacua, without an obvious mechanism to choose
among them, led some authors to advocate the anthropic principle
as a possible resolution for the contrived set of parameters
that seem to govern our world. 

The phenomenological approach to string theory advocates
using the experimental data to study the properties of string
theory. In this respect, it is likely that the type of
backgrounds relevant for the physical observations differ from
the generic backgrounds. For example the extra degrees
of freedom needed to cancel the conformal anomaly will not
appear as continuous dimensions, and hence their a priori 
geometrical interpretation may be misleading. 

The string phenomenology program has by now been pursued for
many years. A particular class of models that exhibit appealing
phenomenological properties are the heterotic string models
in the free fermionic formulation. A key property of this class
of models is the relation of the free fermionic point
in the moduli space and the self--dual point under T--duality.
While the precise relationship needs to be better understood in the 
context of the realistic models,
the self--dual point, being the symmetry point under T--duality,
is the point where the moduli are likely to stabilize. Aside from
being the symmetry point, the self--dual dual point is the one 
where the energy needed to excite the momenta and winding modes
is minimized. It is rather obvious that a function which is invariant
under exchange with its inverse has its minimum, for positive values,
at the self--dual point. 

The second key property of the realistic free fermionic models is their
relation with $Z_2\times Z_2$ orbifold compactification. The special
property of the $Z_2\times Z_2$ orbifold is that it acts on the internal
coordinates as real rather than complex coordinates. This fact has
important bearing on the problem of moduli fixing. In
this context,
recent work \cite{fknr,ron,prep} revealed that the reduction to three 
generations in this class of models is achieved by utilizing 
an asymmetric identification between the left-- and right--moving
internal dimensions. The utilization of the asymmetric shift
has the profound consequence that the untwisted geometrical 
moduli are projected out from the spectrum. Therefore, one 
should no longer regard the internal dimensions as generating
ordinary geometrical objects. The precise nature of the underlying
geometries requires further study and elucidation.

In the string phenomenology approach it is the data that is paving
the way. 
This phenomenological work now opens new vistas that have been
previously unforeseen.
The notion of duality played a pivotal role in the theoretical
developments in particle physics of the past two decades.
Inspired by the phenomenological studies in the context of 
the free fermionic models,
it was proposed here that phase--space duality is the guiding 
property in trying to formulate quantum gravity. 
In this respect T--duality is a key property of string theory.
We can think of T-duality as a phase--space duality
in the sense of exchanging momenta and winding
modes in compact space. We can turn the table around
and say that the key feature of string theory is that
it preserves the phase--space duality in the compact space.
It is further argued that the self--dual points
under phase--space duality are intimately
connected to the choice of the vacuum. The evidence
for this arises from the phenomenological success of the
free fermionic models that are constructed in the vicinity 
of the self--dual point, as well as from the formal derivation
of quantum mechanics from phase--space duality and the
equivalence postulate. 
It will be interesting to explore the notion of self--duality
in the context of the modern nonperturbative duality studies. 
The framework of the Seiberg--Witten theory may provide a
laboratory for such investigations in the sense that we may think
of the curves of marginal stability \cite{cms} as the analog of the 
self--dual points. In this respect the enormous number of
vacua in M--theory may be a mere reflection of the enormity of the
gravitational quantum phase--space.

\section{Acknowledgments}

I would like to thank Ron Donagi, Costas Kounnas, Sander Nooij and 
John Rizos for collaboration and discussions.
I would like to thank the Aspen Center for Theoretical Physics,
The New High Energy Physics Group at Rutgers University, and the
Institute for Advanced Study in Princeton for hospitality in the 
course of this work, 
This  work was supported in part by a PPARC Advanced Fellowship,
by the Royal Society, and by the EU networks, ``Quest for Unification''
and ``Supersymmetry and the Early Universe''.


\vfill\eject

\bigskip
\medskip

\bibliographystyle{unsrt}

\begin{thebibliography}{99}

\bibitem{reviewMtheory} For review and references see {\it e.g.}:\\
M. Duff, hep-th/0403160; hep-th/9912164;\\
B. Ovrut, hep-th/0201032; \\
A. Sen, hep-th/0410103;\\
P. Townsend, hep-th/9612121.

\bibitem{greenschwarz} M.B. Green and J.H. Schwarz, \PLB{149}{1984}{117}.

\bibitem{oxsp} 
\AEF, R. Garavuso and J.M. Isidro, \NPB{641}{2002}{111}; hep-th/0209245;\\
\AEF~ and R. Garavuso, \NPB{659}{2003}{224};\\
\AEF~ and D.J. Clements, \IJMP{19}{2004}{2931}.


\bibitem{fknr} 
A.E. Faraggi, C. Kounnas, S.E.M. Nooij and J. Rizos, hep-th/0311058;
\NPB{695}{2004}{41}.


\bibitem{nilles1} For discussion and references see {\it e.g.}:
H.P. Nilles, hep-th/0410160.

\bibitem{anthropics}  L. Susskind, hep-th/0302219;\\
		      M.R. Douglas, \JHEP{0305}{2003}{046};\\
	T. Banks, M. Dine, E. Gorbatov, \JHEP{0408}{2004}{058};

\bibitem{reviewsofstringtheory} See \eg:\\
M. Green, J. Schwarz and E. Witten, Superstring Theory, 2 vols., Cambridge
University Press, 1987;\\
J. Polchinski, String Theory, 2 vols., Cambridge University Press, 1998;\\
C.V. Johnson, D--branes, Cambridge University Press, 2003.

\bibitem{heterotic} D.J. Gross, J.A. Harvey, E.J. Martinec and R, Rohm,
\PRL{54}{1985}{502}; \NPB{256}{1985}{253}; \NPB{267}{1986}{75}. 

\bibitem{tduality} For review and references see {\it e.g.}:\\
A. Giveon, M. Porrati and E. Rabinovici, \PRT{244}{1994}{77}.

\bibitem{ginsparg} P. Ginsparg, \NPB{295}{1988}{153}

\bibitem{grv} A. Giveon, E. Rabinovici and G. Veneziano,
\NPB{322}{1989}{167}.

\bibitem{egrs} S. Elitzur, E. Gross, E. Rabinovici and N. Seiberg,
\NPB{283}{1987}{413}.

\bibitem{as} See {\it e.g.}: C. Angelantonj and A. Sagnotti, 
			\PRT{371}{2002}{1}.

\bibitem{vwaaf} C. Vafa and E. Witten,  1996
	{\it Nucl.Phys.Proc.Suppl.} {\bf46} (1996) 225;\\
	C. Angelantonj, I. Antoniadis and K. F\"orger, \NPB{555}{1999}{116}.

\bibitem{partitions} \AEF, \PLB{544}{2002}{207}.


\bibitem{hosotani} Y. Hosotani, \PLB{126}{83}{309}; \PLB{129}{83}{193}.

\bibitem{suthree} P. Candelas, G.T. Horowitz, A. Strominger and E. Witten,
                                        \NPB{258}{1985}{46};\\
                M. Dine \etal, \NPB{259}{1985}{549};\\
                B. Greene, K.H. Kirklin, P.J. Miron and G.G. Ross,
                               \NPB{278}{1986}{667}; \NPB{292}{1987}{606}.

\bibitem{twozero} E. Witten, \NPB{268}{1986}{79}; \\
                  J. Distler and S. Kachru, \NPB{413}{1994}{213}.

\bibitem{dhvw} L. Dixon, J. Harvey, C. Vafa and E. Witten,
\NPB{261}{1985}{678}; \NPB{274}{1986}{285}.

\bibitem{narain} K.S. Narain, \PLB{169}{1986}{41};\\
                 K.S. Narain, M.H. Sarmadi and E. Witten, 
				\NPB{279}{1987}{369}.

\bibitem{zthree} L. Ibanez, H.P. Nilles and F. Quevedo,
					\PLB{187}{1987}{25};\\ 
                J.A. Casas, E.K. Katehou, C. Munoz, \NPB{317}{1989}{171};\\
		A. Font, L. Ibanez, F. Quevedo and G. Sierra, 
					\NPB{331}{1990}{421};\\
		J. Giedt, {\it Annals\ Phys.}\/ {\bf 297} (2002) 67.

\bibitem{twoplusone} A. Kagan and S. Samuel, \PLB{284}{1992}{289};\\
	S. Chaudhuri, G. Hockney, J. Lykken, \NPB{469}{1996}{357};\\
	\AEF, hep-th/9511093.


\bibitem{nilles}S. Forste, H.P. Nilles, P.K.S. Vaudrevange and A. Wingerter, 
		hep-th/0406208.

\bibitem{fff}
I.~Antoniadis, C.P.~Bachas and C.~Kounnas, \NPB{289}{1987}{87};\\
H.~Kawai, D.C.~Lewellen and S.H.~Tye, \NPB{288}{1987}{1}.

\bibitem{btd979902} \AEF, ~ hep-ph/9707311;
			    hep-th/9910042;
			    hep-th/0208125;
			    hep-th/0307037;
			    hep-ph/0402029.


\bibitem{nahe} \AEF~ and D.V. Nanopoulos, \PRD{48}{1993}{3288};\\
               \AEF, \IJMP{14}{1999}{1663}.

                
\bibitem{revamp} I. Antoniadis, J. Ellis, J. Hagelin and D.V. Nanopoulos,
				\PLB{231}{1989}{65}.

\bibitem{patisalamstrings} I. Antoniadis, G. Leontaris and J. Rizos,
                                        \PLB{245}{1990}{161}.

\bibitem{fny} \AEF, D.V. Nanopoulos and K. Yuan, \NPB{335}{1990}{347};\\
                \AEF, \PRD{46}{1992}{3204}.


\bibitem{eu} \AEF, \PLB{278}{1992}{131}; \NPB{387}{1992}{239}.


\bibitem{top} \AEF, \PLB{274}{1992}{47};
		    \PLB{377}{1995}{43};
                    \NPB{487}{1996}{55}.


\bibitem{cfn} G.B. Cleaver \etal, \PLB{455}{1999}{135};
                                \IJMP{16}{2001}{425};
                                \NPB{593}{2001}{471};
                                \MODA{15}{2000}{1191};
                                \IJMP{16}{2001}{3565};
                                \NPB{620}{2002}{259}.

\bibitem{lrsstringmodels}
                G.B. Cleaver \etal, \PRD{63}{2001}{066001}; 
                                    \PRD{65}{2002}{106003}.

\bibitem{su421}
                G.B. Cleaver, \AEF~and S.E.M. Nooij, \NPB{672}{2003}{64}. 


\bibitem{ccf}   S. Chang, C. Coriano and \AEF,  \PLB{397}{1997}{76};
                                \NPB{477}{1996}{65};
              J. Elwood and \AEF, \NPB{512}{1998}{42}.

\bibitem{otherrsm} S. Chaudhuri \etal, \NPB{469}{1996}{357};\\
		   G.B. Cleaver \etal, \NPB{525}{1998}{3};
                                \NPB{545}{1998}{47};
                                \PRD{59}{1999}{055005};
                                \PRD{59}{1999}{115003}

\bibitem{cdfd0} F. Abe \etal, \PRL{74}{1995}{2626}; \\
                S. Abachi \etal, \PRL{74}{1995}{2632}.

\bibitem{NRT} \AEF, \NPB{403}{1993}{101}; \NPB{407}{1993}{57}

\bibitem{CKM} I. Antoniadis \etal, \PLB{278}{1992}{257}; \\
\AEF~ and E. Halyo, \PLB{307}{1993}{305}; \NPB{416}{1994}{63};\\
J. Ellis \etal, \PLB{425}{1998}{86}

\bibitem{seesaw} I. Antoniadis, J. Rizos and K. Tamvakis, 
					\PLB{279}{1992}{281};\\
             \AEF~ and E. Halyo, \PLB{307}{1993}{311};\\
             \AEF~ and J.C. Pati, \PLB{400}{1997}{314};\\
	     C. Coriano and \AEF, \PLB{581}{2004}{99}.

\bibitem{gcu} I. Antoniadis, J. Ellis, R. Lacaze and D.V. Nanopoulos, 
			\PLB{268}{1991}{188};\\
              I. Antoniadis, J. Ellis, S. Kelley and D.V. Nanopoulos,
			\PLB{272}{1991}{31};\\
              \AEF, \PLB{302}{1993}{202};\\
              K.R. Dienes and \AEF, \PRL{75}{1995}{2646};
                                           \NPB{457}{1995}{409}.
\bibitem{ps} \AEF, \NPB{428}{1994}{111};
                   \PLB{339}{1994}{223};
                   \PLB{398}{1997}{95};
                   \PLB{499}{2001}{147};   
                   \PLB{520}{2001}{337};\\
        J.C. Pati, \PLB{388}{1996}{532};\\
             J. Ellis \etal, \PLB{419}{1998}{123}.

\bibitem{fp2} I. Antoniadis, J. Ellis, A. Lahanas and D.V. Nanopoulos,
						\PLB{241}{1990}{24};\\
              \AEF~ and E. Halyo, \IJMP{11}{1996}{2357};\\
              \AEF~ and J.C. Pati, \NPB{526}{1998}{21};\\
              \AEF~ and O. Vives, \NPB{641}{2002}{93}.

\bibitem{dedes} A. Dedes and \AEF, \PRD{62}{2000}{016010}.

\bibitem{zp} \AEF~and D.V. Nanopoulos, \MODA{6}{1991}{61};\\
		L.L. Everett, P. Langacker, M. Plumacher and J. Wang,
						\PLB{477}{2000}{233};\\
 		\AEF~and M. Thormeier, \NPB{624}{2002}{163}.

\bibitem{fccp} X.G. Wen and E. Witten, \NPB{261}{1985}{651};\\
	K. Benakli, J. Ellis and D.V. Nanopoulos, \PRD{59}{1999}{047301};\\
        S. Sarkar and R. Toldra, \NPB{621}{2002}{495}.
             
\bibitem{fop} \AEF, K.A. Olive and M. Pospelov, \APJ{13}{2000}{31}.

\bibitem{cfp} C. Coriano and \AEF, \PRD{65}{2002}{075001}; hep-ph/0107304;\\
	C. Corian\`{o}, \AEF~and M. Plumacher, \NPB{614}{2001}{233};\\
	A. Cafarella, C. Coriano and \AEF, \IJMP{19}{2004}{3729};
					hep-ph/0306236.


\bibitem{foc} \AEF, \PLB{326}{1994}{62}.

\bibitem{befnq} P. Berglund \etal, \PLB{433}{1998}{269};
				\IJMP{15}{2002}{1345}.

\bibitem{ron} R. Donagi and \AEF, \NPB{694}{2004}{187}.


\bibitem{kounnas} 
		E.~Kiritsis and C.~Kounnas, \NPB{503}{1997}{117};\\
		A.~Gregori, C.~Kounnas and J.~Rizos, \NPB{549}{1999}{16}.

\bibitem{nooij} S.E.M. Nooij, University of Oxford D.Phil thesis.

\bibitem{prep} A.E. Faraggi, C. Kounnas, S.E.M. Nooij and J. Rizos, 
paper in preparation.

\bibitem{fm} A.E. Faraggi and M. Matone,
			\PRL{78}{1997}{163};
			\PLB{450}{1999}{34};
			\PLB{437}{1998}{369};
			\PLA{249}{1998}{180};
			\PLB{445}{1998}{77};
			\PLB{445}{1999}{357};
			\IJMP{15}{2000}{1869};\\
		G. Bertoldi, A.E. Faraggi and M. Matone, 
				  Class. Quant. Grav. {\bf17} 3965 (2000);\\
		\AEF, hep-th/0003156; hep-th/0312167.



\bibitem{floyd} E.R. Floyd, \PRD{25}{1982}{1547}; \PRD{26}{1982}{1339};
			    \PRD{29}{1984}{1842}; \PRD{34}{1986}{3246};
			    \PLA{214}{1996}{259}; \IJMP{14}{1999}{1111}.


\bibitem{cms} N. Seiberg and E. Witten, \NPB{426}{1994}{19};\\
P.C. Argyres, \AEF~and A.D. Shapere, hep-th/9505190;\\
A. Bilal and F. Ferrari, \NPB{480}{1996}{589};\\
A. Ritz, M.A. Shifman, A.I. Vainshtein and M.B. Voloshin,
\PRD{63}{2001}{065018}.

\end{thebibliography}

\end{document}